\documentclass[10pt,aps,pra,twocolumn, nofootinbib,superscriptaddress]{revtex4-2}
\usepackage{amssymb,amsmath}
\usepackage{color}
\usepackage{hyperref}
\usepackage{listings}
\usepackage{physics}
\usepackage{amsmath}
\usepackage{graphicx}
\usepackage{caption}
\usepackage{subcaption}
\usepackage{bbold}
\usepackage{qcircuit}
\usepackage{multirow}
\usepackage{bm} 
\usepackage[normalem]{ulem}

\usepackage[toc,page]{appendix} 

\usepackage[table,xcdraw,dvipsnames,usenames]{xcolor}

\hypersetup{
    bookmarks=true,         
    unicode=false,          
    pdftoolbar=true,        
    pdfmenubar=true,        
    pdffitwindow=false,     
    pdfstartview={FitH},    
    pdfauthor={},     
    colorlinks=true,       
    linkcolor=blue,          
    citecolor=red,        
}

\definecolor{dkgreen}{rgb}{0,0.6,0}
\definecolor{gray}{rgb}{0.5,0.5,0.5}
\definecolor{mauve}{rgb}{0.58,0,0.82}

\newcommand{\PAOLO}{Scientific Computing Department  STFC-UKRI, Rutherford Appleton Laboratory, Didcot OX11 0QX, United Kingdom}
\newcommand{\LLNL}{Lawrence Livermore National Laboratory, P.O. Box 808, L-414, Livermore, California 94551, USA}
\newcommand{\UNITN}{{Dipartimento di Fisica, University of Trento, via Sommarive 14, I–38123, Povo, Trento, Italy}}
\newcommand{\TIFPA}{INFN-TIFPA Trento Institute of Fundamental Physics and Applications,  Trento, Italy}

\begin{document}
\title{Enhancing Qubit Readout with Autoencoders}

\author{Piero Luchi}
\email{piero.luchi@unitn.it}
\affiliation{\UNITN}
\affiliation{\TIFPA}

\author{Paolo E. Trevisanutto}
\affiliation{\PAOLO}
\author{Alessandro Roggero}
\affiliation{\UNITN}
\affiliation{\TIFPA}

\author{Jonathan L DuBois}
\affiliation{\LLNL}
\author{Yaniv J. Rosen}
\affiliation{\LLNL}
\author{Francesco Turro}
\affiliation{\UNITN}
\affiliation{\TIFPA}

\author{Valentina Amitrano}
\affiliation{\UNITN}
\affiliation{\TIFPA}

\author{Francesco Pederiva}
\affiliation{\UNITN}
\affiliation{\TIFPA}


\begin{abstract}
In addition to the need for stable and precisely controllable qubits, quantum computers take advantage of good readout schemes. Superconducting qubit states can be inferred from the readout signal transmitted through a dispersively coupled resonator. This work proposes a novel readout classification method for superconducting qubits based on a neural network pre-trained with an autoencoder approach. A neural network is pre-trained with qubit readout signals as autoencoders in order to extract relevant features from the data set. Afterward, the pre-trained network inner layer values are used to perform a classification of the inputs in a supervised manner. We demonstrate that this method can enhance classification performance, particularly for short and long time measurements where more traditional methods present lower performance. 
\end{abstract}
\maketitle
\section{Introduction}\label{sec:1}
The construction of a computer exploiting quantum – rather than classical – principles represents a formidable scientific and technological challenge. Nowadays, superconducting quantum processors are reaching outstanding results in simulation~\cite{barends2015digital,wendin2017quantum_info,yan2019quantum_walks,holland2020optimal} and computational power~\cite{arute2019quantum_supremacy}.
However, building a fault-tolerant quantum processor still presents many technical challenges. First of all, it is required the ability to generate high-fidelity gates, exploiting both hardware (e.g. improving the manufacturing process
and the design of available qubits~\cite{nersisyan2019manufacturing,nguyen2019high,huang2020superconducting_review}) and software improvements (e.g. designing precise optimal control protocols~\cite{werschnik2007quantum_control,palao2002quantum_control,kirchhoff2018optimized}). In the second place, one needs the ability to perform a complete quantum error correction protocol~\cite{reed2012QEC,corcoles2015QEC,gong2021QEC}. Finally, it is of primary importance to have a high-fidelity qubit readout measurement to extract information from the device, especially for observables that are very sensitive to it (see e.g.~\cite{roggero2020sqpe} for an extreme case of this). In addition to a careful design of the system parameters~\cite{walter2017rapid,sunada2022fast} or
improvement in fabrication processes extending qubits coherence time~\cite{place2021new,nersisyan2019manufacturing}, readout fidelity can be enhanced through the use of machine-learning techniques. 

The currently most common qubit readout technique is the dispersive readout (in Quantum Electrodynamics, QED, circuit architecture) which couples the qubit to a readout resonator. In this approach, the state of the qubit is determined by measuring the phase and amplitudes of an electromagnetic field transmitted through the resonator~\cite{blais2004cavity,wallraff2005readout,
bianchetti2009readout,gambetta2008readout}.
Hardware, random thermal noise, gate error or qubit decay processes that occur during measurements may reduce the readout fidelity. Machine learning techniques and classification schemes could help to restore a good fidelity by improving the classification precision of the signal to the right state of the qubit. 
Gaussian mixture model \cite{reynolds2009gaussian} is the most commonly used classification method given its ease of use. It exploits parametric modeling of the averaged readout signals probability distribution in terms of a sum of Gaussians to perform a classification of each measurement.
In \cite{magesan2015ML,Seif_2018,lienhard2021ML,
quiroga2021discriminating} the authors {developed and implemented} various classification methods based on neural networks trained on the full dynamics of the measurement, instead of on their averages, obtaining good results.
Another approach is the hidden Markov model proposed in \cite{martinez2020ML}, which allows for a detailed classification of the measurement results and detection of the decay processes that the qubit could undergo during the measurement. These schemes help to improve the accuracy of the classification of the qubit readout measurements.

In this work, we propose a novel semi-unsupervised machine learning classification method based on autoencoder pre-training applied to the heterodyne readout signal of a superconducting qubit \cite{blais2004cavity,wu2020high}.
Autoencoders are a type of artificial neural network designed to encode a set of data efficiently by learning how to regenerate them from a synthetic encoded representation \cite{bishop2006pattern,bengio2012unsupervised}. The encoding process automatically isolates the most relevant and representative features of the input dataset, i.e. those features which allow for the most faithful reconstruction of input data while neglecting noise and non-relevant details~\cite{pasa2014pretraining,ong2014pretraning}.
Hence, the main idea of this work is to exploit this characteristic of autoencoders and perform the data classification not on the readout signals or on their time average, but on their encoded representation produced by autoencoder training.
The model consists of two sections. The first is composed of an autoencoder trained to reconstruct the qubit readout signals dataset. The second section is a two-layer feed-forward neural network trained to classify the encoded representation of the measurement signals. 
We demonstrate that this method can enhance the state classification of readout signals, especially for short readout times where other more traditional methods have worse performance and, in general, shows a more stable performance for a broad range of measurement time lengths. We remark on the fact that the most significant improvement occurs with a combination of hardware and software improvements, as obtained by the authors in Ref. \cite{chen2022transmon}. In this paper, the focus will be only on software improvement on present machines.

The paper is organized as follows. In Sec. \ref{sec:1}, qubit setup and readout, a review of machine learning models of interest, as well as our proposed method, are presented. In Sec. \ref{sec:2}, the method is tested on two study cases, based on real data, and the classification results together with considerations of the method's applicability are presented. Finally, in Sec. \ref{sec:3}, conclusions are drawn.

\section{Methods}\label{sec:2}
\subsection{Qubit readout}\label{subsec:2A}
We consider a transmon-type qubit coupled to a detuned resonator (i.e. a quantum harmonic oscillator) in the context of a strong projective dispersive measurement scheme \cite{wallraff2005readout,bianchetti2009readout}. Our device is a qutrit with frequency $\omega_{01}=2\pi \times 3.44$ GHz, anharmonicity $\alpha=-2\pi \times 208$ MHz, and a Rabi frequency of $5$ MHz. It has a relaxation time $T_1=220 \ \mu$s and a coherence time $T_2=20 \ \mu$s. The device is based on the work by Place \textit{et al.} \cite{place2021new}. A $100$ ns constant $\pi-$pulse of frequency $\omega_{01}$ is used to rotate the qutrit from $\ket{0}$ to $\ket{1}$, and another $100$ ns $\pi-$pulse to rotate the qutrit from $\ket{1}$ to $\ket{2}$. The coupling energy between the qubit and the cavity, (before the dispersive approximation) is $2\pi \times 107 $ MHz.
The cavity has a frequency $\omega_r=2\pi \times 7.24$ Ghz. The measurement pulse is a constant pulse with $\omega_r$ frequency of arbitrary duration. A Purcell filter is present in the device.

Due to the qubit interaction, the readout resonator undergoes a frequency shift whose value depends on the qubit state. This dependency can be exploited to perform measurements of the qubit state in the dispersive regime i.e. when the detuning of qubit and resonator is large relative to their mutual coupling strength \cite{kohler2018dispersive}. Once the resonator is irradiated with a specific microwave pulse, the registered transmitted signal will incorporate different amplitude and phase shifts based on the qubit's state. The demodulation procedure can extract such information from the signal, discriminating between qubit states.

\begin{figure}[t]
    \centering
    \includegraphics[width=0.5\textwidth]{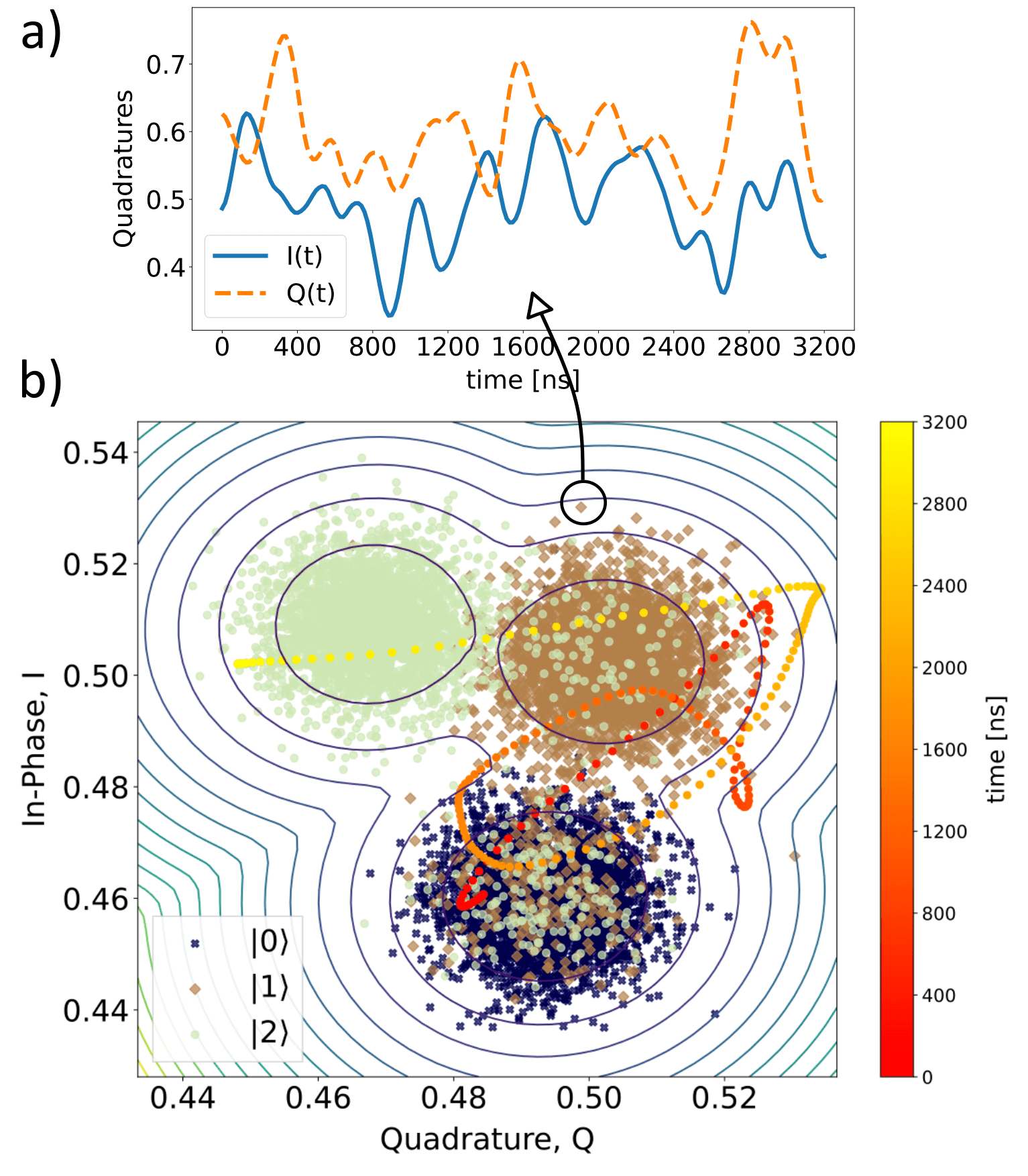}
    \caption{Pictorial representation of qubit readout data.
    \emph{Panel a} Example of in-phase, I(t), and quadrature, Q(t), components of heterodyned signal of a single shot obtained via sliced demodulation (as described in Sec.\ref{subsec:2A}). The average of these signals is a single point in the I-Q plane below. 
    \emph{Panel b} Example of the whole dataset. Each point is the time average of a measurement represented in the I-Q plane for qubit states 0,1, and 2. The lines represent the 2D Gaussian contour plot (see Sec. \ref{subsec:2B1}) for the 3 Gaussian distribution. The dotted red-yellow line is an example of a measurement signal represented in the I-Q plane. The colors represent the time evolution (in nanoseconds).}
    \label{fig:traj}
\end{figure}

{Our setup consists of a superconducting qubit controlled by the Quantum Orchestration Platform (QOP) programming environment (Q.M Technologies Ltd.) through the QUA programming language based on Python \cite{QUA}.} 
In this setup, the measurement pulse is sent into the readout resonator. In interacting with the system, this signal is modulated by the resonator's response. The output signal is then filtered, amplified, and down-converted to an intermediate frequency $\omega_{IF}=\omega_r-\omega_{LO}$ through a signal mixer, with $\omega_{LO}$ the frequency of the local oscillator (an electronic component needed by the mixer to change signals frequency).
Finally, it has to be demodulated to extract information about the qubit state that the readout signal acquired in the interaction.

Formally, the demodulation is an integral of the signal multiplied by a sinusoidal function:

\begin{eqnarray}
I&=&\frac{2}{T_m}\int_{0}^{T_m}r(\tau)\cos(\omega_{IF}\tau)d\tau\\
Q&=&-\frac{2}{T_m}\int_{0}^{T_m}r(\tau)\sin(\omega_{IF}\tau) d \tau,
\end{eqnarray}  

\noindent where the readout signal is denoted by $r(\tau)$ and $T_m$ is the integration time.

In the usual approach, complete demodulation is performed by integrating over time intervals $T_m$, obtaining a single value for the I and Q components for each qubit readout signal. In this way, each measurement can be represented as a point in the I-Q plane. Thanks to the qubit state-dependent frequency shift, these points will accumulate in different zones of the I-Q plane. An example is displayed in panel (\emph{b}) of Fig.~\ref{fig:traj} where the points for a three-level qutrit are reported.

\begin{figure}[ht]
  \centering
  \begin{subfigure}[t]{0.45\textwidth}
    \includegraphics[width=\textwidth]{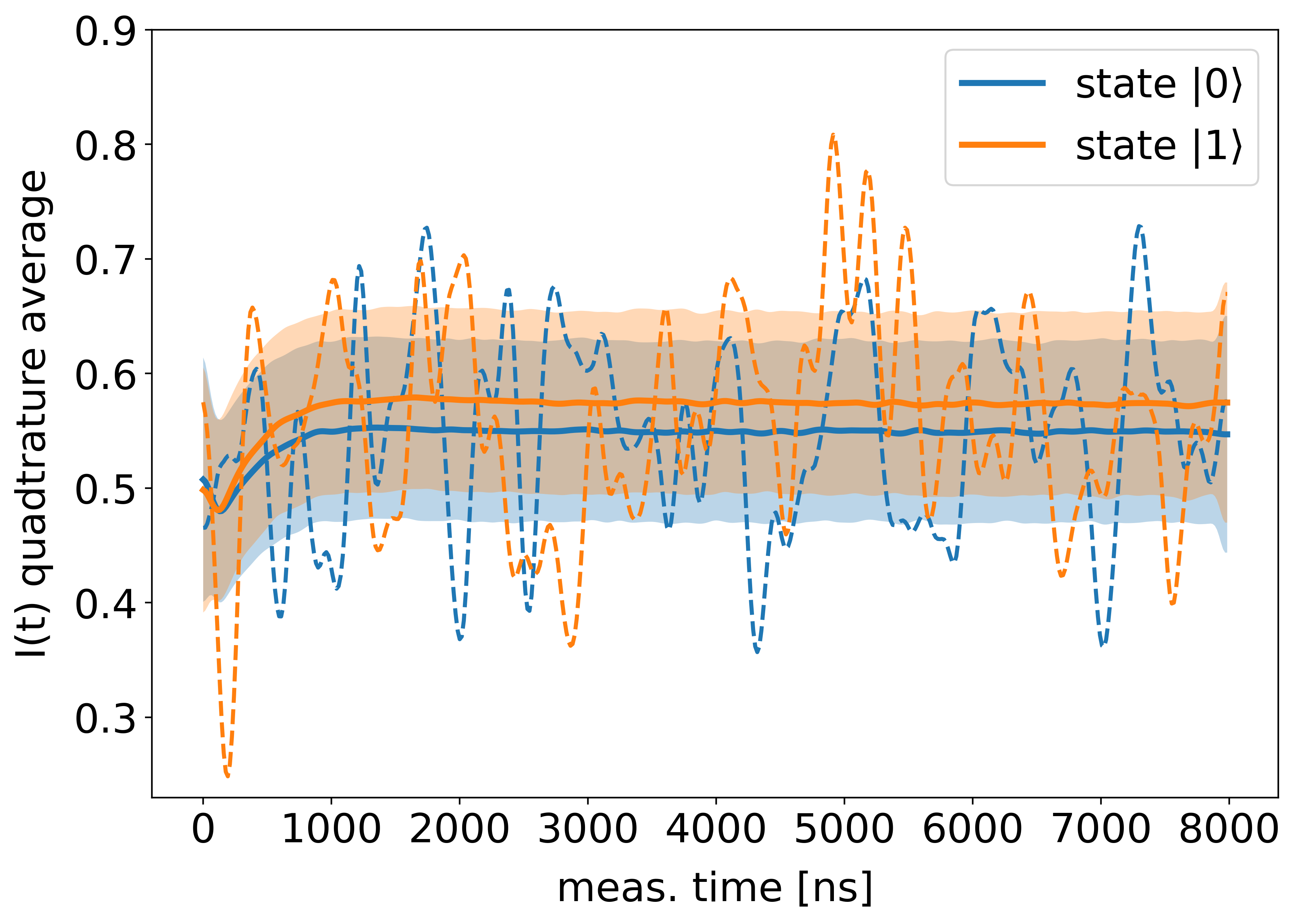}
    \caption{In-phase, $I(t)$ signals}
    \label{fig:I_average}
  \end{subfigure}
  \hfill
  \begin{subfigure}[t]{0.45\textwidth}
    \includegraphics[width=\textwidth]{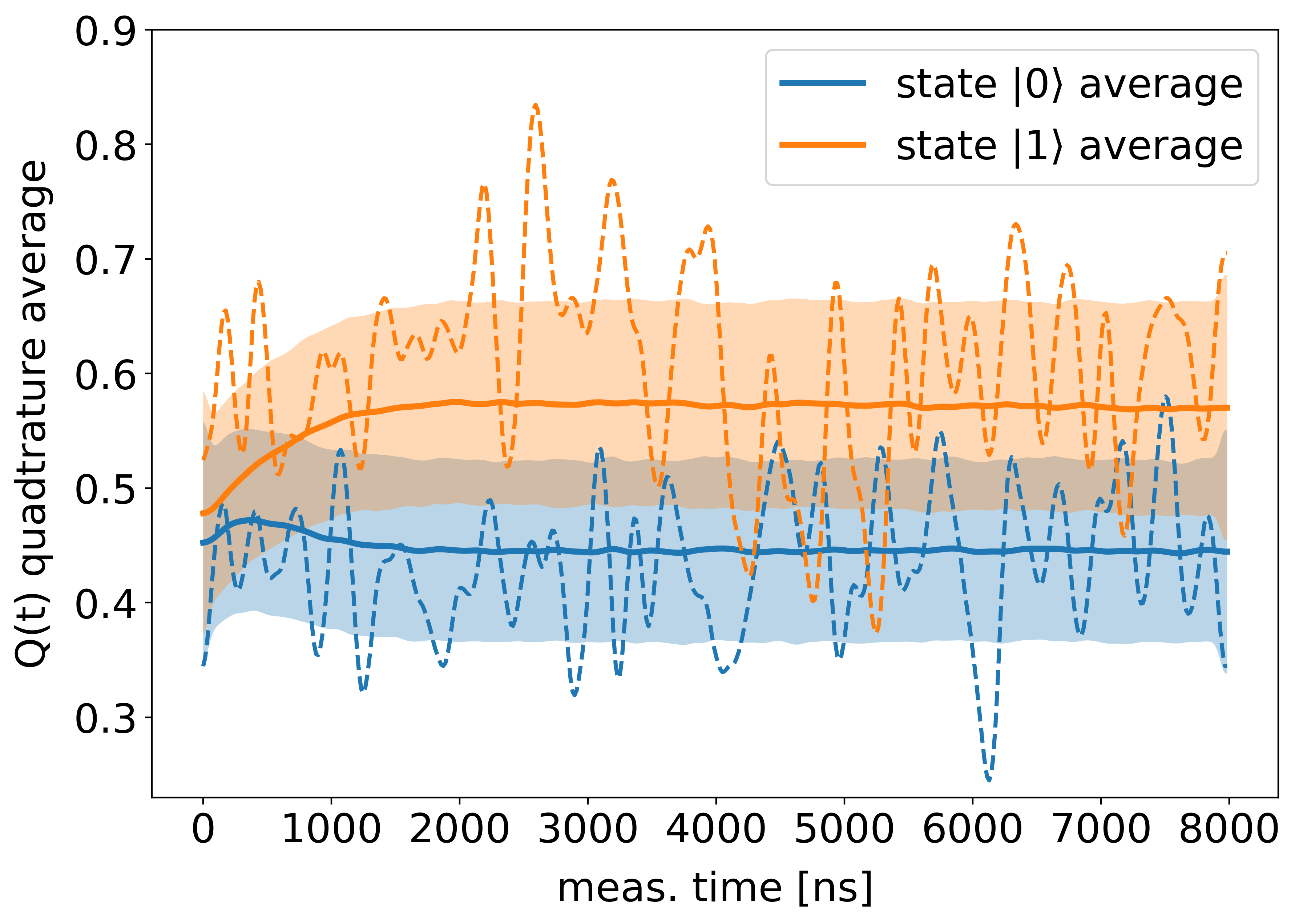}
    \caption{Quadrature, $Q(t)$ singals}
    \label{fig:Q_average}
  \end{subfigure}
  \caption{Average readout trajectories for state $\ket{0}$ and $\ket{1}$ in both quadrature. Solid lines represent the mean of all trajectories in the data set for state $\ket{0}$ (blue) and state $\ket{1}$ (orange). The shaded regions represent the standard deviation of the average for each timestep. The dashed line instead represents an example of a single trajectory.}
  \label{fig:traj_averages}
\end{figure}

However, an alternative approach can be employed, the so-called \emph{sliced demodulation}, which consists in dividing the  time interval $[0, T_m]$ into $N$ subintervals and performing the demodulation separately on each chunk of the signal, namely: 

\begin{eqnarray}
I(t)&=&\frac{2}{\Delta t }\int_{t}^{t+\Delta t}r(\tau) \cos(\omega_{IF}\tau)d\tau\\
Q(t)&=&-\frac{2}{\Delta t}\int_{t}^{t+\Delta t}r(\tau)\sin(\omega_{IF}\tau)d\tau\;,
\end{eqnarray} 
with $\Delta t = T_m/C$ the subinterval length and $C$ the number of intervals. 
In this way, we obtain two time series, $I(t)$ and $Q(t)$ for each measurement. {In Fig. \ref{fig:traj}, panel (\emph{a}), the $I(t)$ and $Q(t)$ signals of a single readout signal are represented as examples. Averaging these signals we obtain a single point in the I-Q plane as represented in the panel  (\emph{b}). The red-yellow line in panel  (\emph{b}) represents the $I(t)$ and $Q(t)$ signals plotted together as a trajectory (state-path trajectory). The color gradient represents time.
In Fig. \ref{fig:traj_averages}, instead, the average signals $\left< I(t) \right>  \rangle$ and $\left< Q(t)\right>$ (solid lines) are reported together with the standard deviation for each timestep (shaded range). The same graph also shows individual readout signals (dashed lines). As can be seen, the machine's noise is high as the standard deviation zones heavily overlap. However, one of the aims of this work is to show how the proposed method can deal with this noise and improve, in any case, the classification of the measures.}

In principle, the sliced demodulation should retain information that otherwise is lost in the averaging process of the complete demodulation. This information will be exploited in this work to increase state detection accuracy.
Usually, in full demodulation, the readout accuracy is adjusted and maximized by tuning the readout length, i.e. the demodulation integration time $T_m$. The aim is to obtain clouds of points (as in Fig.~\ref{fig:traj}) with a distribution that is as Gaussian as possible to use the Gaussian Mixture Model to perform the classification (see. Sec.~\ref{subsec:2B}). In fact, short integration times produce poorly distinguishable states, while for long times, the qubit states tend to decay during the measurement, which produces a non-Gaussian data distribution and, again, low classification accuracy.
In contrast to full demodulation, sliced demodulation retains more information about the qubit state measurements and, in principle, allows for increased accuracy of the state classification. Moreover, as will be observed in this work, it reduces the dependence of the classification result on $T_m$ since the data do not need to be Gaussian distributed.\\

\subsection{Machine Learning Models}\label{subsec:2B}
We briefly review the three machine learning algorithms used in this work.

\subsubsection{Gaussian Mixture Model}
\label{subsec:2B1}
\emph{Gaussian mixture models} (GMM) approximate distribution of data (in this case, the clouds of mean demodulation points in panel (\emph{b}) of Fig.\ref{fig:traj}) as a weighted superposition of Gaussian distributions \cite{reynolds2009gaussian}. The GMM models the distribution by {adjusting the Gaussian parameters through a maximum likelihood estimation} over the dataset of points in the I-Q plane. A new point is attributed to one of the classes based on the probability that it belongs to one of the three Gaussians of the GMM.

\subsubsection{Feed-forward Neural Network}
\emph{Feed-forward neural networks} (FFNN) are the simplest class of neural networks. Trained over a labeled dataset, they are capable of classifying new inputs. Formally the neural network implements a closed-form parametrized function, $N_\phi$, which maps input in $\mathcal{X}  \subseteq  \mathbf{R}^m$ into a space $\mathcal{Y}  \subseteq  \mathbf{R}^n$ which encodes in some way the information on the classes the inputs are divided into. The inputs are the full qubits readout signals.
An optimal classification of data is obtained by adjusting the parameters $\phi$ making use of optimization algorithms. This is obtained by minimizing some type of loss function $l$ between the correct label $\bm y^i$ of input $\bm x^i$ and the neural network predicted label $\hat{\bm y}_i = N_\phi(\bm x^i)$, namely:

\begin{eqnarray}\label{eq:loss_NN}
\min_\theta \sum_i l\left(\bm y^i,N_\phi(\bm x^i)\right)\;.
\end{eqnarray}

\noindent This optimization is commonly carried out by making use of the well-known \emph{back-propagation} algorithm~\cite{bishop2006pattern,yu2002backprop}.

\subsubsection{Autoencoders}
\textit{Autoencoders} are neural networks designed to learn, via unsupervised learning procedures, efficient encoding of data~\cite{Goodfellow2016auto,bengio2012unsupervised,shrestha2019autoen}. This encoding is achieved by adjusting the network's weights and biases to regenerate the input data. It is composed of a first part, the encoder, which learns to map the input data into a lower dimensional representation (the latent space), ignoring insignificant features or noise, and a second part, the decoder, that, conversely, is trained to reconstruct the original input from the low dimensional encoding in the latent space.
Autoencoders perform dimensionality reduction and feature learning.

Mathematically, the autoencoder is a model composed of two closed-form parametrized functions, the encoder $f_{\theta^e}$ and the decoder $g_{\theta^d}$. The parameters $\theta = [\theta^e , \theta^d]$ need to be optimized to perform the correct inputs reconstruction. These functions are defined as:

\begin{eqnarray}
\nonumber f_{\theta^e}&:& \mathcal{X} \to \mathcal{L} \\
\nonumber g_{\theta^d}&:& \mathcal{L} \to \mathcal{X}\;. 
\end{eqnarray}

\noindent The function $f_{\theta^e}$ takes an input $\bm x^i \in \mathcal{X}  \subseteq \mathbf{R}^m$ from the data-set $ \left \{ \bm x^1, \bm x^2, ... \right \}$ and maps it into the feature-vector $\bm h^i \in \mathcal{L} \subseteq \mathbf{R}^p $ with $p<m$ i.e. $\bm h^i=f_{\theta^e}(\bm x^i)$. Conversely, the decoder function, $g_{\theta^d}$ maps the feature-vector $\bm h^i$ 
back into the input space, giving a reconstruction $\tilde{\bm x}^i$ of the input $\bm x^i$.

\noindent The parameters $\theta$ of the autoencoder are optimized such that the model minimizes the reconstruction error $l(\bm x, \tilde{\bm x})$, i.e. a measure of the discrepancy of the reconstructed input from the original one. The general minimization problem is, therefore:

\begin{eqnarray}\label{eq:loss_autoenc}
\min_\theta \sum_i l\left ( \bm x^i,g_{\theta^d}(f_{\theta^e}(\bm x^i))\right ).
\end{eqnarray}
Again, this is optimized with the already mentioned \emph{back-propagation} algorithm \cite{bishop2006pattern,yu2002backprop}.

\subsection{Model: Neural Network with Autoencoder type Pre-training}\label{subsec:2C}
In this work, we propose a classification model based on a neural network with an autoencoder pre-training which we denote "\emph{PreTraNN}". It is composed of two sections.

The first section consists of an encoder $f_{\theta^e}$ whose parameters $\theta^e$ are pre-trained in advance as an autoencoder over the input dataset. The encoder consists of two layers with $L_1$ and $L_2$ neurons and a third layer, the latent layer, with $L_H$ neurons. The decoder $g_{\theta^d}$ necessary for the pre-training has the same structure as the encoder but is in reverse order. Given a input of dimension $d$, we always set  $L_1=\frac{3}{4}d$, $L_2=\frac{2}{4}d $ and $L_H=\frac{1}{4}d$. The activation functions are the \emph{sigmoid} for the first layer of the encoder (and the last layer of the decoder) and the \emph{tanh} function for all the internal layers.
 The choice of internal layer size is explained in Appendix \ref{subsec:appA} while the complete specifications of the autoencoder are reported in Appendix \ref{sec:appendixB}.

The second section is a feed-forward neural network, $N_\phi$, dependent on a set of parameters $\phi$ which works as a classifier taking as inputs the feature-vector of the encoder and, as outputs, the exact labels of the readout signals. It is composed of two hidden layers with $ L_{N_1}$ and $L_{N_2}$ neurons, respectively, and an output layer with a number of neurons equal to the number of data classes. Given $d$ the dimension of the input, we set $L_{N_1}=2d$ and $L_{N_2}=d$. The activation functions are \emph{tanh} for the internal layers and the \emph{softmax} for the last layer, commonly employed for classification purposes.

The assignment of the label $\bm y^i$  to a qubit readout signal $\bm x^i(t)$ works as follows:

\begin{enumerate}
\item The discrete signal $\bm x^i$ is flattened by stacking the I and Q components in a single one-dimensional vector, i.e $\bm X^i=[\bm x^i_I, \bm x^i_Q]$ so it can be plugged into the neural network.
\item The input $\bm X^i$ is transformed in the feature-vector $\bm h^i$ via the encoder function, i.e. $\bm h^i=f_{\theta^e}(\bm X^i)$.
\item The feature-vector $\bm h^i$ is plugged into the feed-forward neural network $N_{\phi}$ to be assigned to one out of the three classes. Formally, $N_{\phi}(\bm h^i)= \hat{\bm y}^i$ where $\hat{\bm y}^i$ is the predicted label for the input $\bm X^i$.
\end{enumerate} 

\begin{figure}[t]
    \centering
    \includegraphics[width=1.0\linewidth]{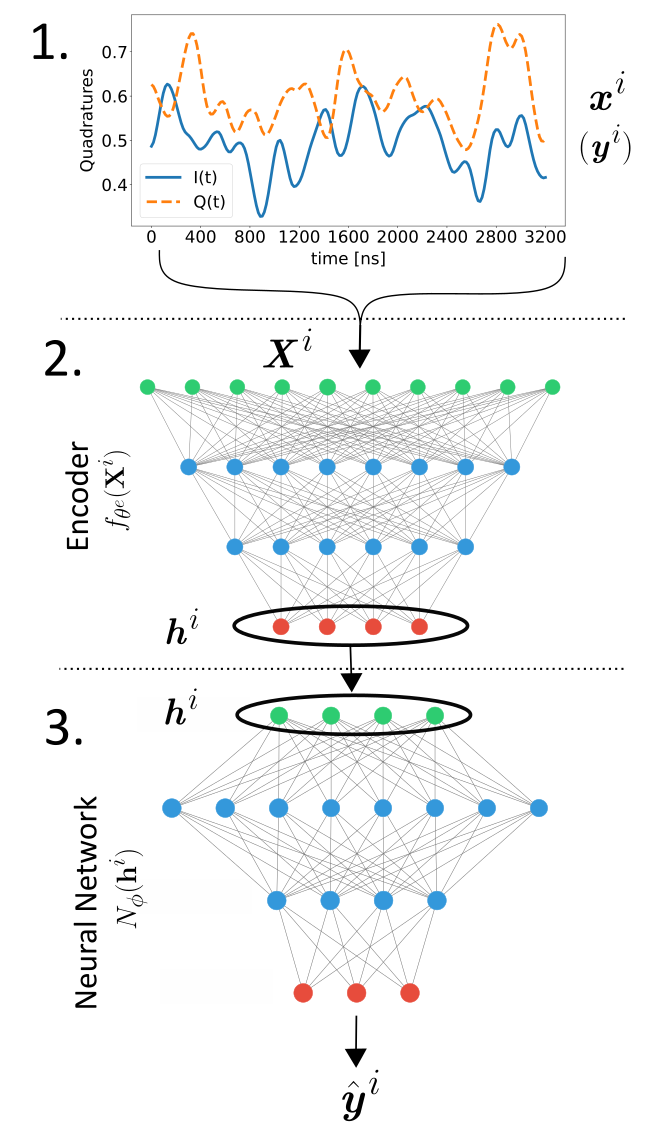}
    \caption{Pictorial representation of the working principle and the architecture of the PreTraNN method described in Sec. \ref{subsec:2C}. \emph{Section 1}: Example of the measurement signal $\bm x(t)$ we want to classify with PreTraNN. \emph{Section 2}: The input $\bm x(t)^i$ is flattened to obtain $\bm X^i$, plugged into the encoder, previously trained as an autoencoder, and transformed into its encoded representation $\bm h^i$. \emph{Section 3}:  The latent layer of the encoder,$\bm h^i$ is passed into a feed-forward neural network trained to assign the label $\hat{\bm y}^i$.}
    \label{fig:pictorial_explanation_prediction}
\end{figure}

\noindent A pictorial representation of the PreTraNN classification working principle is displayed in Fig. \ref{fig:pictorial_explanation_prediction}.

\begin{figure}[ht]
    \centering
    \includegraphics[width=1.0\linewidth]{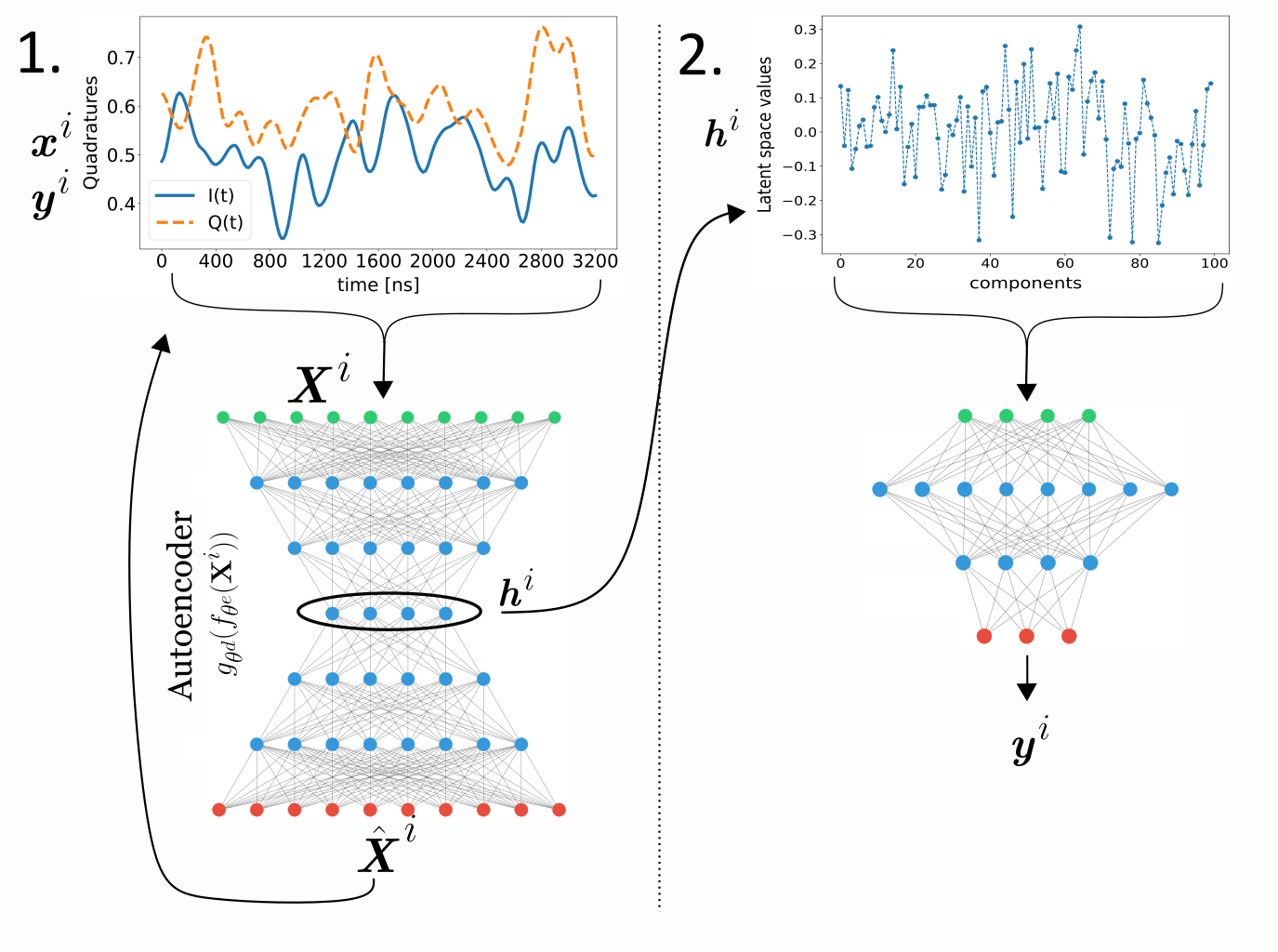}
    \caption{Pictorial representation of PreTraNN training described in Sec. \ref{subsec:2C}. \emph{Section 1}: The autoencoder is trained to reconstruct the measurement signals. This should train the network to extract the relevant features from each temporal chunk. \emph{Section 2}: After the training, the decoder part of the network is removed, and the encoded representation of data (represented in the plot at the top right) is used as the train input dataset for the second section of the PreTraNN model which is trained to classify them into the correct class $\bm y^i$}
    \label{fig:pictorial_explanation_training}
\end{figure}

\subsubsection{Training}
The training is performed separately for the two sections that compose the PreTraNN model. 

The autoencoder is trained first. The dataset is composed by inputs $\bm x^i$ with $i = 1,2,...,M$,
representing the 2D trajectories in the I-Q plane. The neural network architecture requires a one-dimensional vector input so $\bm x^i$ need to be flattened, stacking the I and the Q components in a single one-dimensional vector. So we compose a new dataset of $\bm X^i= [\bm x_I^i \  \bm x_Q^i] $. The parameters $\theta=[\theta^e , \theta^d]$ of the autoencoder $A_{\theta}(\bm x^i)=g_{\theta^d}(f_{\theta^e}(\bm X^i))$ are trained by minimizing  Eq.~\eqref{eq:loss_autoenc} where we choose as loss function $l$ the mean square error 
\begin{eqnarray}
\label{eq:auto_loss}
l=\frac{1}{d}\sum_{t=1}^{d}\left(X^i[t] - \hat{X}^i[t]\right)^2,
\end{eqnarray} 
with $d$ the length of the input data $\bm X^i$ and $ \hat{\bm X}^i=A_{\theta}(\bm X^i)$ the reconstructed input. 

In a second step, the neural network $N_{\phi}$ is trained taking as inputs the feature-vectors $\bm h^i$ of the encoder $f_{\theta^e}$ and, as output, the real labels  $\bm y^i$ of the corresponding $\bm x^i(t)$. The optimal network's parameters $\phi$ are obtained by minimizing Eq.~\eqref{eq:loss_NN} where the loss function $l$ is chosen to be the cross entropy loss function, widely used in classification.

A pictorial explanation of the PreTraNN training procedure is depicted in Fig. \ref{fig:pictorial_explanation_training} while a complete specification of the autoencoder structure is reported in Appendix \ref{sec:appendixB}.

{Note that although the use of the term 'pre-training', PreTraNN is not a pre-trained model in the general sense. We do not use a bulk neural network pre-trained on a vast quantity of data, attaching to it new layers which are then trained on our specific classification problem. In PreTraNN model, we take as  “pre-trained neural network” the encoder part of an autoencoder that was previously trained over our specific readout data. }

\subsection{Benchmark Methods}\label{subsec:2D}
We compare the result of the proposed PreTraNN model with two state-of-the-art methods introduced above: the Gaussian mixture model (GMM) and a simple feed-forward neural network (FFNN).

The GMM is trained directly on I-Q points, averages of the readout signal.

The FFNN is, instead, trained over the readout signals dataset, taking as input the flattened vectors $\bm X^i=[\bm x_I^i, \bm x_Q^i]$ and, as outputs, their labels $\bm y^i$. The architecture of the FFNN consists of two inner layers of dimension $L_{FF_1}=2d$ and $L_{FF_2}=d$, with d the input dimension, and an output layer.  The activation functions are the \emph{tanh} for the internal layer and the \emph{softmax} for the output layer. The structure of the FFNN is the same as the second section of the PreTraNN. The only difference is that while the PreTraNN neural network takes as input the readout signal encoded in the latent space, the $\bm h^i$ vector, the FFNN takes directly the signals $\bm X^i$.

\subsection{Metrics}\label{subsec:2D}

To measure the accuracy of the classification systems, we utilize the "classification accuracy", i.e. the probability that each signal is attributed to the correct label (i.e. the correct state of the qubit). This classification is obtained as a percentage of correctly attributed signals out of their total number (for each state). The global accuracy is the average of the accuracies of each state.

\subsection{Datasets}\label{subsec:2F}
As already mentioned, two versions of the same dataset are used in this work. They will be now more clearly defined.

We collect heterodyned readout signals for each qubit state. Each measurement is obtained by preparing the device in states (e.g. $ \ket{0} \mathrm{or} \ket{1}$) and then by measuring it immediately, storing the obtained signals. The selection of the time windows $\Delta t$ for the sliced demodulation requires careful consideration. The demodulation time-step $\Delta t$ should span an integer number of periods of the readout signal to avoid imprecise demodulation. The frequency of the readout signal is $\omega_{IF}=60$ MHz, so its period is $1/\omega_{IF}\approx 16 \ \mathrm{ns}$. 
For this reason, in this work, we took a time window $\Delta t=16$ ns. Hence, each readout signals $\bm x^i(t)$ has a point every $16$ ns.
The length of the measurement, $T_m$ is also an essential parameter. Here we choose to consider measurements of increasing length starting from $800$ ns up to $8000$ ns, corresponding to discrete signals whose number of elements spans from $50$ to $500$, to study the efficiency of the classification methods in different configurations. 
{The collection of $I(t)$ and $Q(t)$ signals are then smoothed with a window smoothing algorithm with a Hanning window of 50 timestep length to remove some noise.}

Each measurement is, therefore, a two-dimensional $\bm x^i(t)=[I^i(t), Q^i(t)]$ trajectory that, flattened to form the $\bm X^i$ inputs (See Sec.~\ref{subsec:2C}), will form the dataset for the PreTraNN and FFNN. 
The dataset for the GMM, on the other hand, is obtained by time-averaging each $\bm x^i(t)$ measurement so as to obtain two values that can be represented in I-Q space (an example of which is shown in panel (\emph{b}) Fig.~\ref{fig:traj}). The dataset is then shuffled and split into train and test datasets in a 75\% - 25\% proportion.
The size of the dataset impacts the accuracy of the method and needs some consideration to avoid under-fitting or unnecessarily long training times. Such considerations are drawn in Appendix~\ref{subsec:appB}.

{It should be mentioned that data preparation is not error-free. In fact, it may happen that the expected state ($\ket{0}$,$\ket{1}$ or $\ket{2}$) is not actually prepared due to control errors or environmental coupling. The $\ket{0}$ state is initialized with an active reset procedure. This procedure works by performing a short measurement on the qubit and applying a $\pi-$ pulse to it if the state is $\ket{1}$ or $\ket{2}$ to push it back to the ground state.  However, there will be a residual thermal population to deal with. An estimate of this quantity, for the two-state case, is given by the confusion matrices in Fig. \ref{fig:confusion_matrices_01}, which quantify the percentage of $\ket{1}$ labeled measurements that are actually $\ket{0}$.
Due to these errors, the classification will not be 100\% accurate even with the model proposed in this paper because the dataset suffers from this inaccuracy.}

{We also emphasize that the readout data are all from the same device. Although it may be interesting to study a multi-device classification system, in general, different devices may show differences in the average behavior of the readout trajectories, due to design and control differences. This obviously makes training more challenging and could bias the results.}

\section{Results}\label{sec:3}
The purpose of this work is to demonstrate how the feature extraction capability of the autoencoder helps improve the effectiveness of qubit readout. So, specifically, how the PreTraNN method performs better than other commonly used methods for readout, namely GMM and a simple FFNN. In this section, PreTraNN and the benchmark methods are compared in terms of classification accuracy and their overall performance is studied.

In addition, to deepen the analysis, the application of the models is extended to two readout configurations. The first is the readout of the usual two-level qubit and the second is the readout of a three-level qutrit.  This analysis will give an idea of the good scalability of PreTraNN for  multiple levels readout. 


\begin{figure}[t]
  \centering
  \begin{subfigure}{\linewidth}
    \centering
    \includegraphics[width=1.0\linewidth]{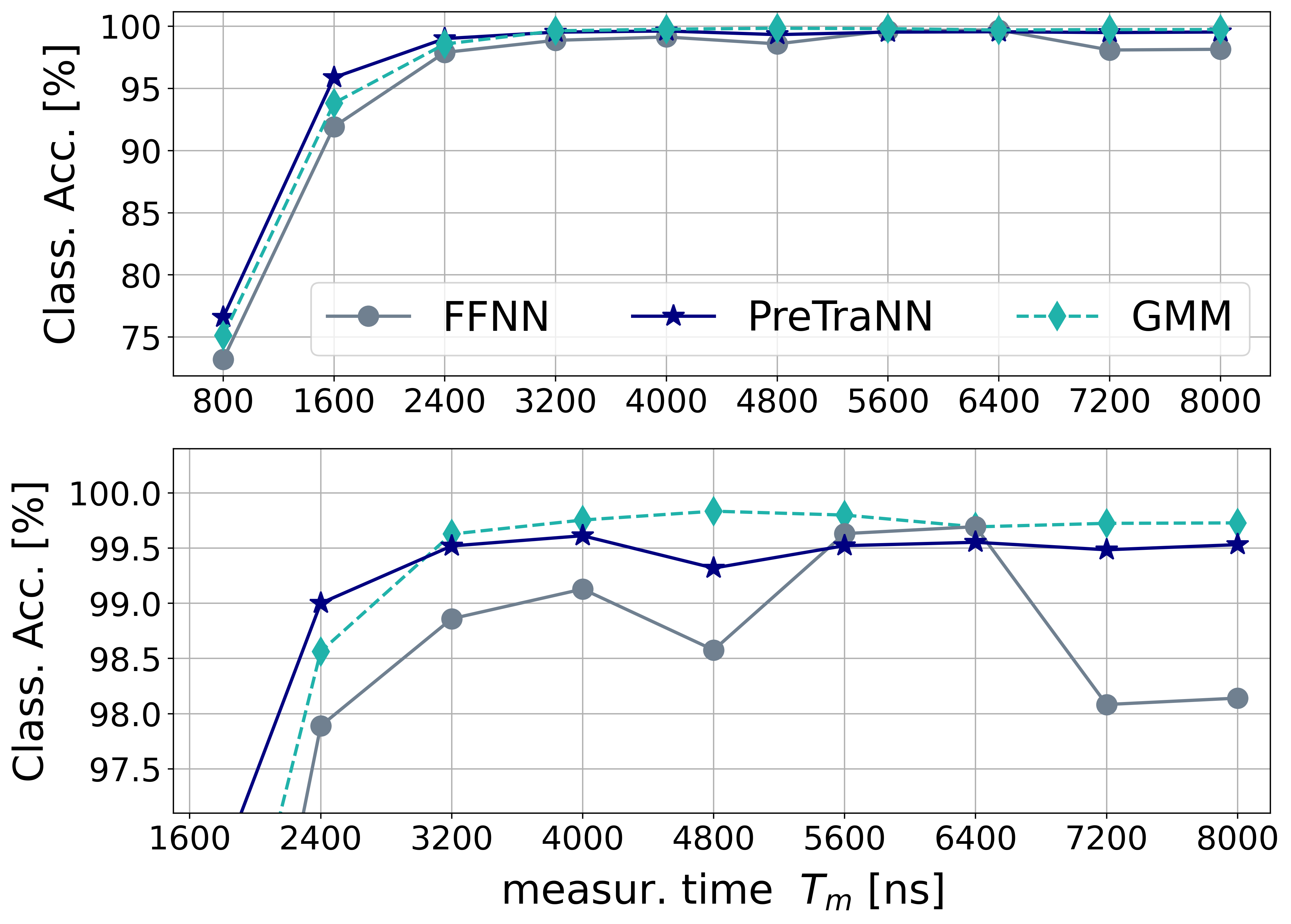}
    \caption{\emph{Upper panel}: Classification accuracy for state $\ket{0}$ by the three methods as a function of the measurement time. \emph{Lower panel}: a zoom on the 2400-8000 ns part of the plot.}
	\label{fig:0}
\end{subfigure}
\begin{subfigure}{\linewidth}
    \centering
    \includegraphics[width=1.0\linewidth]{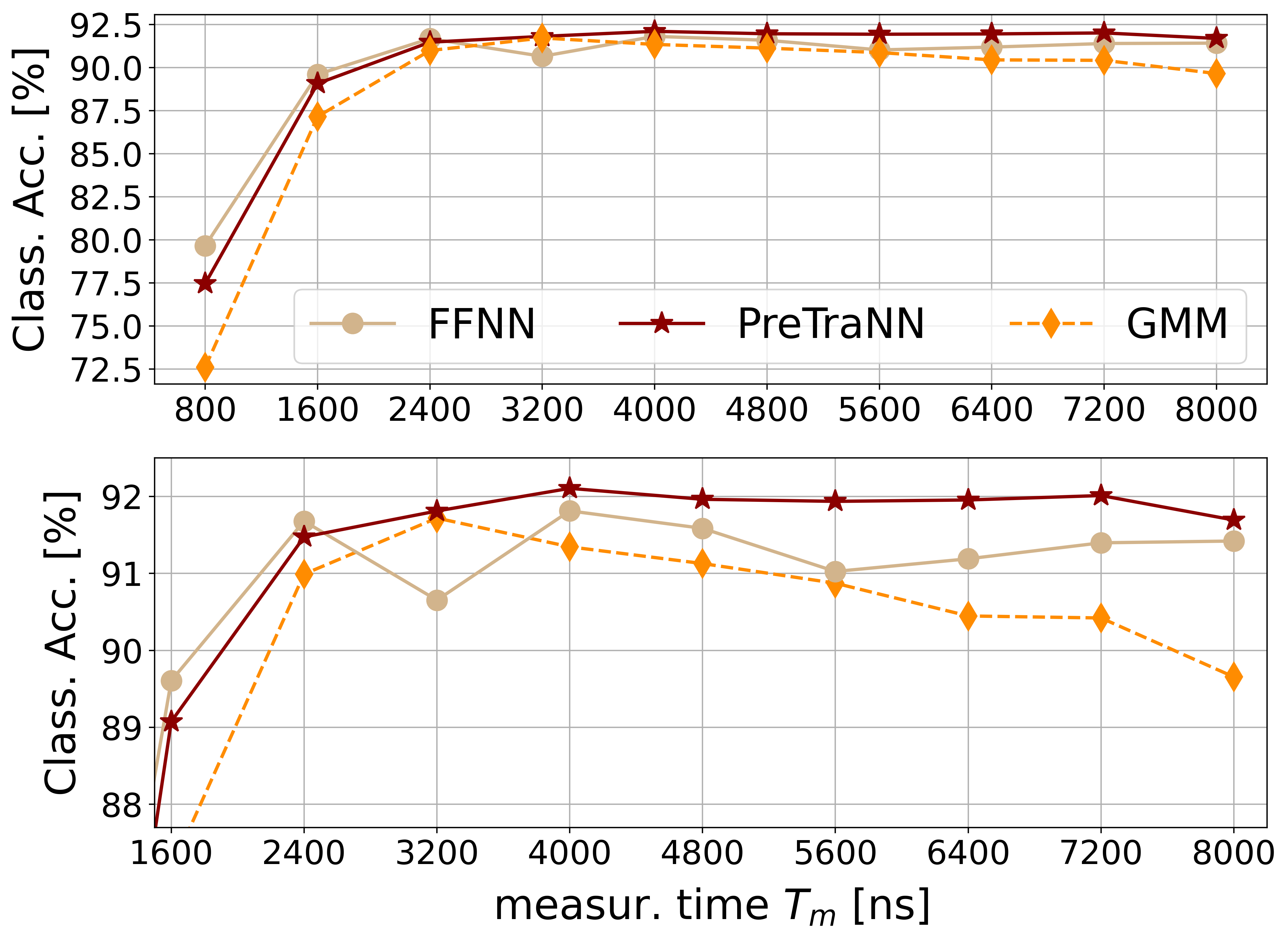}
    \caption{\emph{Upper panel}: Classification accuracy for state $\ket{1}$ of the three methods as a function of the measurement time. \emph{Lower panel}: a zoom on the 2400-8000 ns part of the plot.}
    \label{fig:1}
  \end{subfigure} 
  \caption{Classification accuracy comparison, for state $\ket{0}$ and $\ket{1}$ separately, between Gaussian Mixture Model (GMM), the simple feed-forward neural network (FFNN) and the PreTraNN method . The readout time $T_m$ spans from $800$ ns to $8000$ ns. }  
\label{fig:accuracy}
\end{figure}  

\begin{figure}[h]
    \centering
    \includegraphics[width=1.0\linewidth]{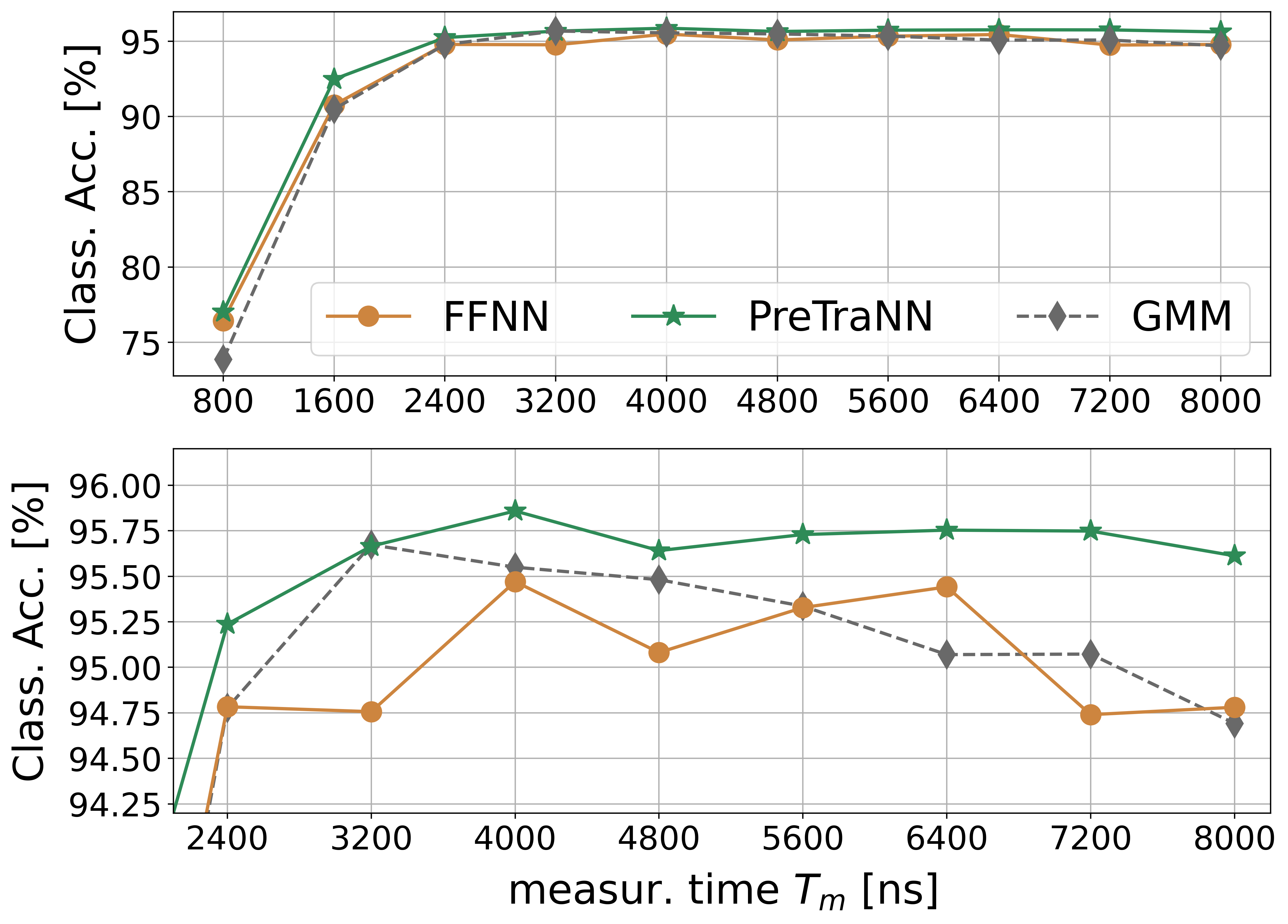}
    \caption{Global classification accuracy between state $\ket{0}$ and $\ket{1}$ for increasing measurement time $T_m$. The accuracy obtained with PreTraNN method is higher (or at most equal) to the ones obtained with GMM and FFNN.}
\label{fig:accuracy_tot}
\end{figure}

\begin{figure*}[ht]
\centering
\includegraphics[scale=0.45]{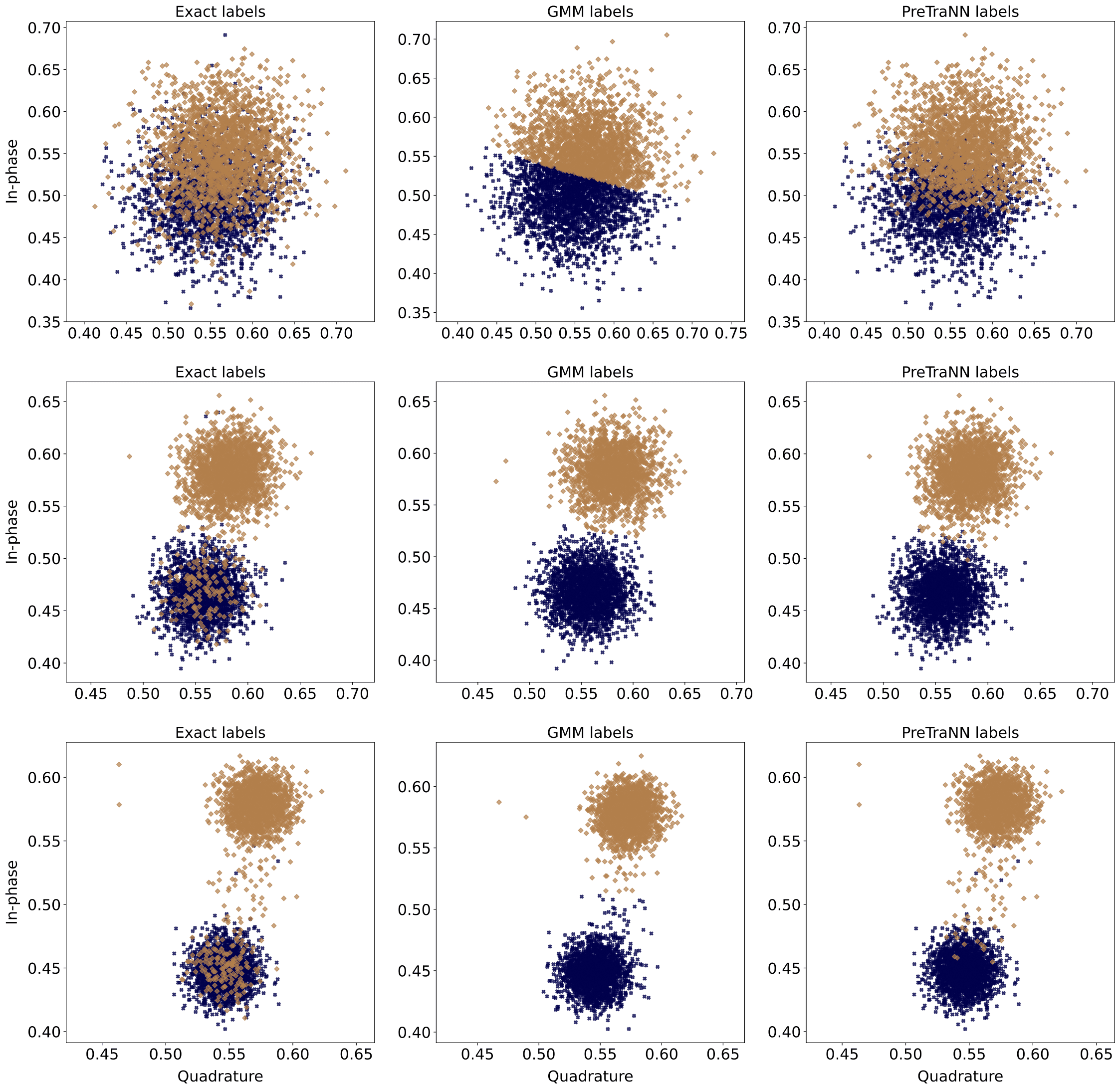}
\caption{Pictorial representation of the dataset with exact, GMM's and PreTraNN's labeling. Each point is the time average of the I(t) and Q(t) signals. The actual label, i.e. the prepared state, is represented in the first column. The GMM and PreTraNN methods labels are represented in the second and third columns.}
\label{fig:clouds_accuracy_meas_time}
\end{figure*}

\subsection{Two-state qubit readout}\label{subsec:3A}
In this case, the qubit is prepared and immediately measured in state $\ket{0}$ and $\ket{1}$. The dataset consists of 16000 readout signals (8000 for each state) and it is split into train and test subsets in 75\%/25\% proportions. Consideration on the choice of the dataset are drawn in Appendix~\ref{subsec:appA}.
The PreTraNN, FFNN  and GMM setup is the one defined in Sec.~\ref{subsec:2C} and Sec.~\ref{subsec:2D}.

\subsubsection{Classification accuracy}
We start by showing our results for the classification accuracy of the three methods for increasing measurement length $T_m$ to compare their performance in different cases. All experiments are carried out in the configuration defined in Sec.~\ref{sec:2} and every experiment is computed 10 times and averaged. We report the state classification accuracy for each state separately in Fig.~\ref{fig:accuracy} and the global classification between state $\ket{0}$ and $\ket{1}$ in Fig.~\ref{fig:accuracy_tot}.

We start by considering Fig. \ref{fig:accuracy}. 
In the upper figure, the classification accuracy of $\ket{0}$ state for the three models as a function of measurement length is shown, in the lower figure, the same information is reported but for $\ket{1}$ state. First of all, it can be noted that, for short measurements, all models deteriorate their performance. This behavior should be attributed to the fact that, for short measurement times, the data distributions heavily overlap, preventing all methods, and especially GMM, from fitting them appropriately with two Gaussians (see Fig.~\ref{fig:clouds_accuracy_meas_time} for an illustrative example). 
For middle and long measurement times, instead, the GMM performs, respectively, better and worse for state $\ket{0}$ and state $\ket{1}$ than the other two methods. Moreover, the state $\ket{0}$ classification accuracy remains high and stable for long measurements, while that of state $\ket{1}$ presents a descending trend at longer times. This behavior has a simple explanation: {the qubit excited state (e.g. the $\ket{1}$ state) has leakage to the ground state (the $\ket{0}$ state) at a much higher rate than the opposite direction}. As a consequence, there is an asymmetry in the data points distributions. This results in states prepared as $\ket{1}$ to be spotted on state $\ket{0}$ distribution due to the decay process, while the reverse is much more unlikely. Therefore the GMM, fitting the distribution with two Gaussians, can not handle this asymmetry performing very differently in the two cases. The number of signals decaying during the measurement procedure increase with the measurement time and, in fact, the accuracy of state $\ket{1}$ drops for long times. Instead, FFNN has a fluctuating trend, and it performs often worse than GMM. We can speculate that this behavior derives from the fact that, for very large inputs, the training is more difficult, and a simple FFNN does not converge adequately. This suggests that FFNN is not completely adequate for this purpose. On the contrary, the PreTraNN method shows very stable behavior for both states even for long measurement times. It not only uses all the "history" of the measurements but also exploits the feature extraction of the autoencoder.

In Fig.~\ref{fig:accuracy_tot}, the global discrimination accuracy between state $\ket{0}$ and $\ket{1}$ is reported. It is obtained averaging the accuracies of $\ket{0}$ and $\ket{1}$ states. In this global case, the PreTraNN method outperforms the GMM and the FFNN methods for every measurement time (except for a measurement time of 3200 ns where GMM's and PreTraNN's accuracies coincide). The considerations of the previous case also apply here.

It can also be noted that GMM accuracy has a global maximum at $3200$ ns. As mentioned before, for the GMM to work well, the distribution of I-Q points for each qubit state must be as "Gaussian" and distinguishable as possible. It happens that, for short measurement times, the points distributions overlap since the qubit-resonator response is still in a transient state, while, for long times, decaying processes come into play which makes the distribution skewed. Therefore, we can deduce that the length of $3200$ ns produces the least overlapping distributions that allow the GMM to reach the greatest accuracy. This measurement time is therefore the one that should be set for the readout in case of GMM use. The PreTraNN method makes the need for this adjustment less strict since it works well for a larger interval of the experimental parameters $T_m$. In general, it can be seen that, in PreTraNN method, the classification accuracy is only increasing or constant. As a consequence, the trimming is faster and easier since the need of finding the maximum accuracy is removed.

We want also to stress that in other works, such as Ref. \cite{magesan2015ML}, the readout accuracy may be greater than the one reported here. As described above, the machine used for this work has a certain level of error in preparing state $\ket{1}$. This, however, is of secondary importance since the purpose of the present work was not to present new hardware over-performing the current state-of-the-art one, but only to propose a method to improve readout in the present machines. Thus, interest was primarily focused on improving the performance of a given machine from the software point of view. 
\\
The classification obtained with PreTraNN, not only improves the classification accuracy but also better reproduces the actual distribution of data. In Fig.~\ref{fig:clouds_accuracy_meas_time} the comparison of the GMM and PreTraNN labeling result on data with different readout times is reported. The labeling for the FFNN is similar to the
PreTraNN one, so it was omitted for clarity purposes. The first column shows data with the actual labels (represented by colors) as they were prepared in the quantum device. The second and third columns, instead, represent the same data but labeled according to GMM and PreTraNN, respectively. The same analysis is performed for short, medium and long times (rows of the figure).
As anticipated, we conclude again that the GMM misses the classification for short times, dividing simply in half the overlapping distributions, while the PreTraNN provides a considerably more realistic and accurate classification. The two distributions of points overlapping can now be spotted again. 

The exact labels show the asymmetry in the data distribution due to the decay of the excited state: many $\ket{1}$-labeled points lay in $\ket{0}$ distribution.
The comparison between the labels highlights that there are many points belonging to state $\ket{1}$ that even PreTraNN fails to recognize. Probably many of those points result from the imperfect calibration of the $\pi$-pulse used to prepare the state $\ket{1}$ on the machine.

\begin{figure*}[ht]
\centering
\includegraphics[scale=0.40]{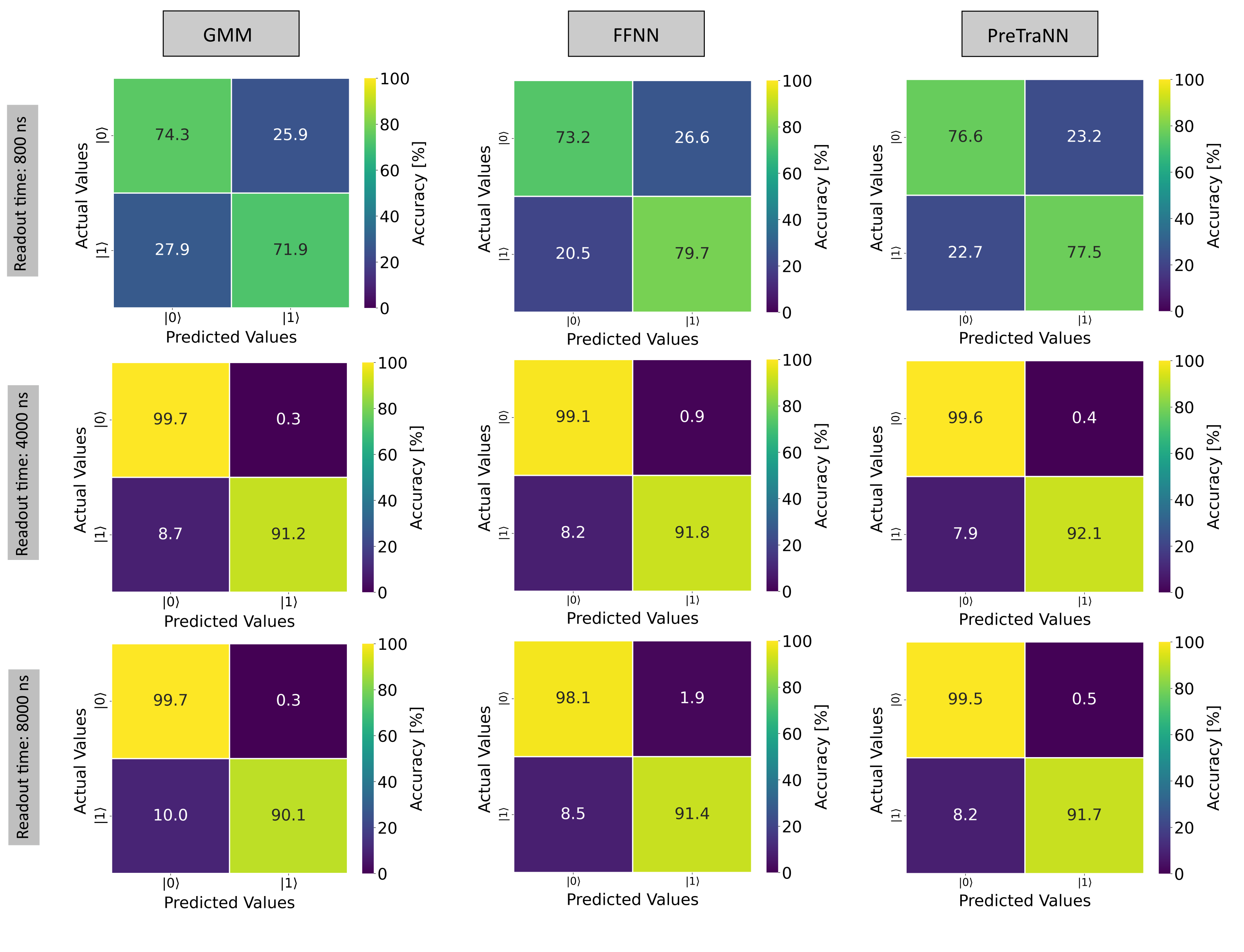}
\caption{Confusion matrices for classification between states $\ket{0}$ and $\ket{1}$ for the three methods for short, medium and long readout times. } 
\label{fig:confusion_matrices_01}
\end{figure*}

Another important measure to take into account is the confusion matrix, which helps to visualize the classification performances of the three methods in comparison. In Fig.~\ref{fig:confusion_matrices_01} are reported the confusion matrices for the three methods in three different measurement length setups. Each row reports the confusion matrices of the three models for a specific measurement length. Clearly, the best confusion matrices are those obtained for long times and with PreTraNN model.

\subsubsection{Computational cost and scaling}
The higher structural complexity of the PreTraNN architecture means training and classification times longer than GMM. In the following, we report the results together with some consideration on the scaling of the method. 

\begin{figure}[t]
\centering
\includegraphics[width=1\linewidth]{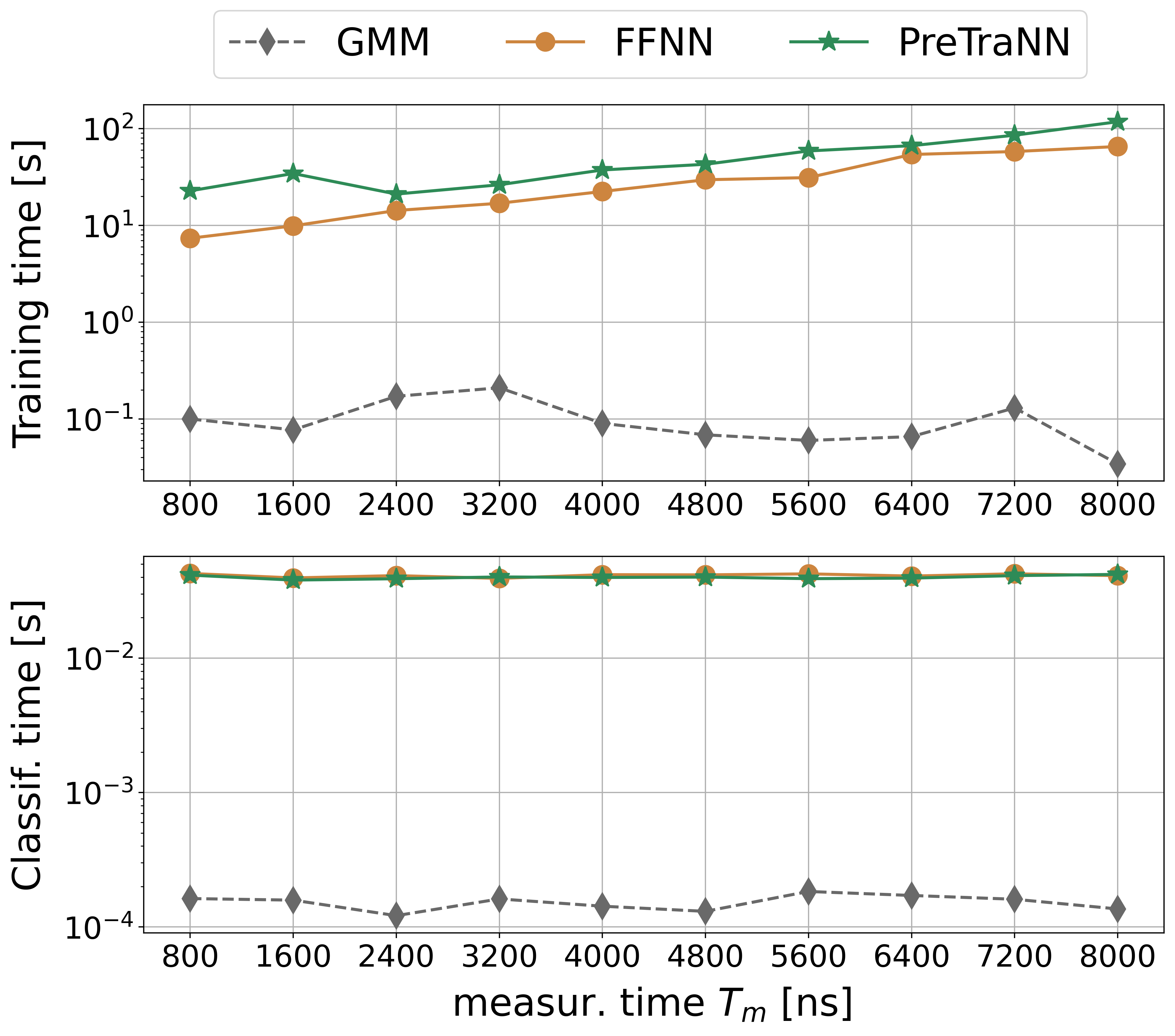}
\caption{Training and classification times for GMM, FFNN and PreTraNN methods. The times are reported in seconds for a middle-range laptop computer. \emph{Upper panel}: Training time in function of the measurement time (i.e. the length of the inputs). \emph{Lower panel}: Classification time. The average time is  0.00013  seconds for GMM, 0.039 seconds for FNN and 0.037 seconds for PreTraNN. }
\label{fig:times_01}
\end{figure}

The training for every neural network is performed with the "early stopping" approach to avoid over- or under-fitting. Instead of fixing the number of epochs, the training is stopped when the accuracy of the model does not increase for two epochs in a row. In Fig.~\ref{fig:times_01} are reported the results. The upper table shows the training time of each model with respect to readout length $T_m$ for a 16000 elements dataset, the lower table, instead, represents the average time for a single input classification for each method. In both cases, the times are represented in logarithmic scale to better spot trends. Times are reported in seconds and refer to a mid-range laptop computer with 4 cores and 8 GB of RAM.

Considering the training time, it can be noted that the PreTraNN method takes a significantly longer time than the parameter estimation for GMM (from 2 to 3 orders of magnitude) but not much more than the FFNN, despite the two training stages of the PreTraNN. As one might expect, the training time of non-GMM methods increases as the inputs measurement time increases. In fact, long measurement times correspond to wider neural networks and, therefore, longer optimizations.
 
From the classification time point of view, we see that the times of the PreTraNN to label a single data (0.039 and 0.042 seconds, respectively) are almost equal and much longer than GMM's (0.00013 seconds). Moreover, for each method, the classification time does not depend on the measurement length.

It is important to specify that the classification time of an inputs batch of size $S$ is not $S$ times the classification time of a single input. We report the actual classification times as a function of batch size in Tab. \ref{tab:1}.

\begin{table}[]
\begin{tabular}{l|l|l|l|l|l}
input batch size & 1  & 100  & 10000 \\ \hline
\begin{tabular}[c]{@{}l@{}}Classif. time \\ PreTraNN {[}s{]}\end{tabular} & 0.04200 & 0.04300  & 0.22400 \\ \hline
\begin{tabular}[c]{@{}l@{}}Classif. time\\  GMM {[}s{]}\end{tabular} & 0.00012 & 0.00013 &  0.00043
\end{tabular}
\caption{Classification times for PreTraNN and GMM as a function of inputs batch size. Every reported time is the result of an average of 100 experiments. The FFNN method is not reported because its behavior follows PreTraNN's.  \label{tab:1}}
\end{table}

Based on this data, some considerations can be made. 
First of all, we can assert that the training for PreTraNN and FFNN remains easily manageable by any computer, even for the longest measurement times. In fact, the training times, although much larger than the GMM, remain very small in absolute value. In general, the training process is not a problem since is done in advance.

On the classification time side, instead, more careful considerations must be made. If only an offline classification is needed, there are no stringent time constraints, and the model could be considered fast enough for some applications. If one instead needs a real-time or online readout on the machine, the classification times must be below the qubit lifetime. Since state-of-the-art superconducting transmon qubits have a lifetime of 200-500 microseconds \cite{kjaergaard2020superconducting,
huang2020superconducting_review}, in principle, we want a classification time that is well below these values, possibly on the order of tens or hundreds of nanoseconds. For this goal, neither the GMM nor PreTraNN has, under the conditions used in this work, the necessary characteristics. Of course, with the use of more powerful computers, the classification time can be reduced by a few orders of magnitude. Moreover, an FPGA or an ASIC implementation could improve even more the efficiency of the classification step or also improve the training process by implementing it in an online way. See Ref. \cite{westby2021fpga,saric2020fpga,
jewajinda2010fpga,
gandhare2019survey}.

To summarize, the ability to perform short-time measurement classification (with higher accuracy) is of great interest in quantum computing. The proposed approach allows for good accuracy for short measurements compared to GMM. This can be exploited for real-time control systems, e.g., quantum orchestration platforms, leading to measurement speed-up or reducing computational time in error correction routines. Attention must be paid to the classification speed of the system. At least partially, however, the longer time required to perform classification can be compensated for by shorter measurements (as little as 1000 ns) than those of the GMM (4000 ns) while achieving the same classification accuracy. The PreTraNN performs well regardless of readout time, allowing one to potentially skip the readout time $T_m$ trimming. Moreover, this method can be utilized for usual two-level qubits or, conversely, extended to arbitrary numbers of levels or qubits with slight modifications in its structure and simply using different datasets. All this considered, the proposed method offers a promising approach to exploit short measurements that disturb the device as little as possible with less computational effort.

%
%
%

\subsection{Three-state qutrit}\label{subsec:3B}
In this case study, we exploit the possibility of accessing the higher quantum levels of superconducting qubits. We prepare and measure the qubit in $\ket{0}$,$\ket{1}$ and $\ket{2}$ state and store the obtained data. The whole dataset consists of 24000 elements (8000 for each state) divided into 75\% train data and 25\% test data. Again, for consideration of how the dataset is chosen, see. Sec. \ref{subsec:2C}. The architecture of the models is the same as in the previous case (and as defined in Sec.~\ref{subsec:2C} and Sec.~\ref{subsec:2F} ). The only difference between the two cases is the number of classes in the dataset. This allows to show the good scaling properties of the model.

\begin{figure}[ht]
    \centering
    \includegraphics[width=1.0\linewidth]{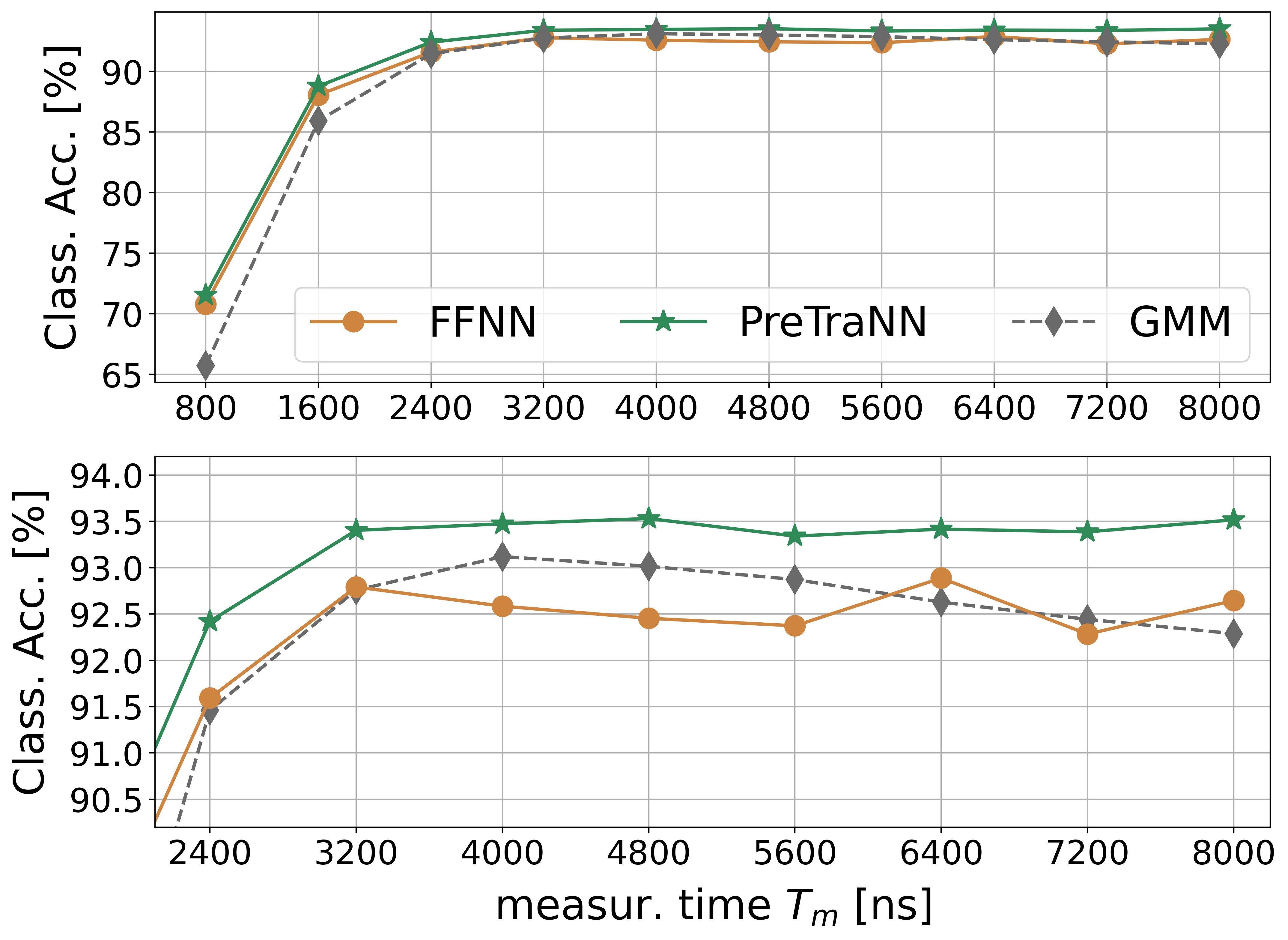}
    \caption{Global classification accuracy for  $\ket{0}$,$\ket{1}$ and $\ket{2}$ states classification for a qutrit. }
    \label{fig:accuracy_tot_012}
\end{figure}

%

\subsubsection{Classification accuracy}
In this paragraph, the global classification accuracy is reported and discussed.


In Fig.~\ref{fig:accuracy_tot_012} we present results for the global accuracy. The PreTraNN method achieves better classification performance for every measurement time. Again the GMM accuracy presents an increasing and decreasing trend with a maximum located at 4000 ns, while the FFNN, notwithstanding a reduction in the fluctuating trend, obtains a lower classification accuracy  than the other two methods possibly due to training difficulties for high dimensional datasets. The PreTraNN instead presents a stable accuracy as a function of measurement time. See Appendix \ref{sec:appendixD} for further details on the state-by-state classification accuracy.

\begin{figure}[ht]
\centering
\includegraphics[width=1.0\linewidth]{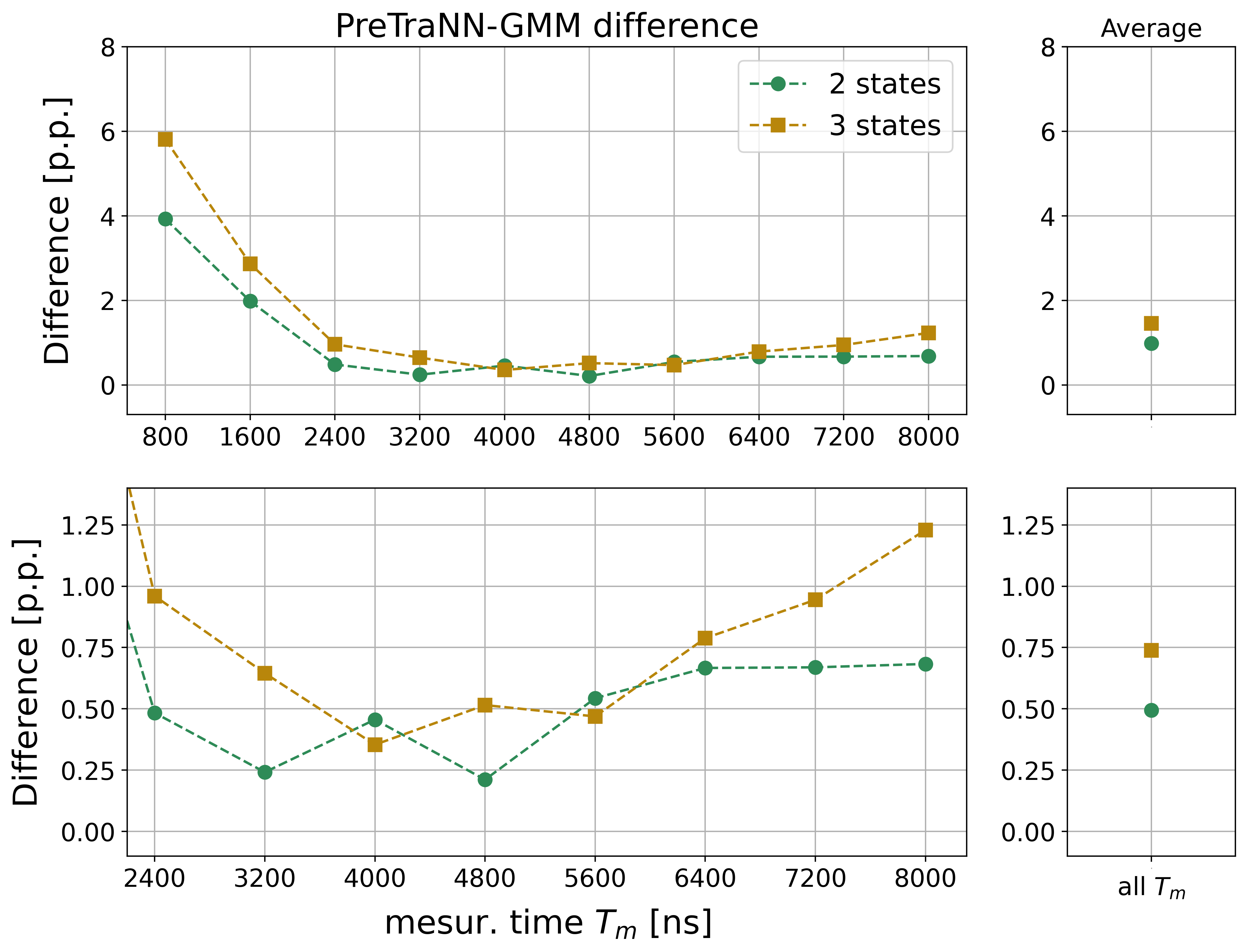}
\caption{Difference in percentage points [p.p.] between the accuracy of PreTraNN and GMM for the qubit and qutrit cases for different measurement time $T_m$. The lower panel reports the analysis only for medium-long times. The small panels on the right show the average of all values of the respective plot on the left. }
\label{fig:difference_pp_GMM}
\end{figure}

\begin{figure}[ht]
\centering
\includegraphics[width=1.0\linewidth]{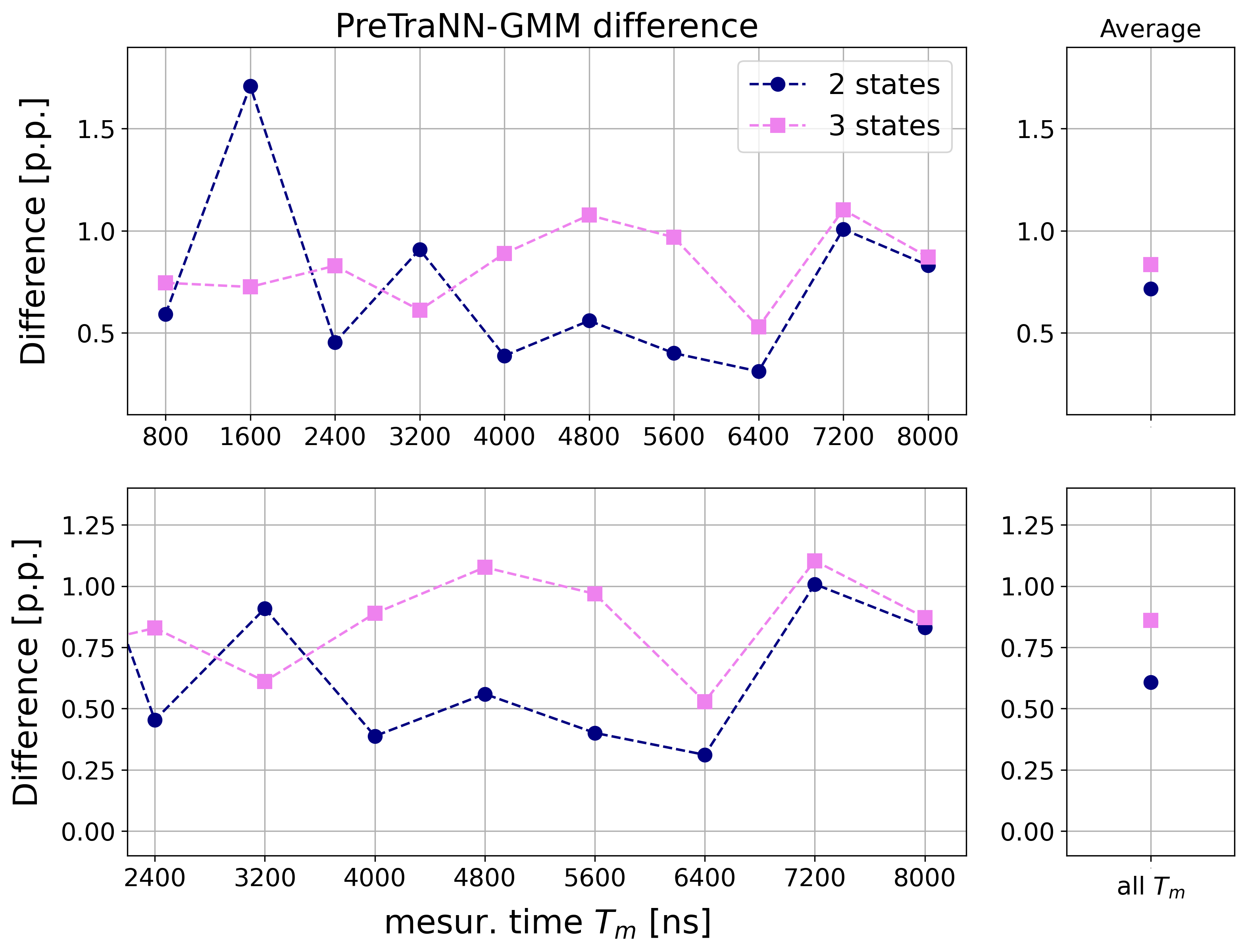}
\caption{Difference in percentage points [p.p.] between the accuracy of PreTraNN and FFNN for the 2 or 3 qubit state case. The lower panel reports the analysis only for medium-long times. The small panels on the right show the average value of the respective plot on the left.}
\label{fig:difference_pp_FFNN}
\end{figure}

We can also study the performances of PreTraNN as a function of the number of qudit levels. This will give us an idea of how the method scales with the number of points clouds. To achieve this, we compute the difference in percentage points (p.p.) between the global classification accuracy of the PreTraNN and that of the other methods. In Fig.~\ref{fig:difference_pp_GMM} the difference in (p.p.) between the PreTraNN global accuracy and GMM's global accuracy for the two and three-level cases for every measurement time $T_m$ is reported. The lower panel zooms on the middle and long times range. On the right panels, the average values for all $T_m$ are highlighted. An increasing value of this difference, as the levels of the system increase, suggest a possible increasing advantage in using the PreTraNN method for increasing system levels. In this case, we see that this trend can be clearly seen. Fig.~\ref{fig:difference_pp_FFNN}, instead, reports the same calculation referred to FFNN method. Here, too, the trend is clear for both the whole set of measurement times and the medium-long range.

This analysis suggests that there is a marginal increase in the effectiveness of PreTraNN compared to the other two methods as the classes of the dataset increase (i.e., as the dataset complexity rises). In other words, the difference in the global classification accuracy between PreTraNN and GMM or between PreTraNN and FFNN is bigger, on average, in the case of the three classes dataset, corresponding to qutrit readout data.

This analysis, although limited to 2 and 3 classes problem, suggests that the PreTraNN method should scale well as the qudit dimension increase. We can assume that it also scales well with the number of qubits since it also reduces to a multiclass dataset, but further analysis to better characterize the performance is needed. 

Furthermore, PreTraNN requires only minimal structural modifications for different qudit dimensions. One only needs to adjust the number of output nodes in the last stage of the network and use an appropriate dataset with a different number of classes. 
While the training times rise due to the increased dataset size (training time grows linearly with the dataset dimension), the classification time remains the same as the previous 2-state case.

\section{Conclusion}\label{sec:5}

This work demonstrates that a feed-forward neural network with autoencoder pre-training allows for a robust qubit readout classification scheme with high accuracy and low dependence on the experimental device feature values. It allows for a consistent classification performance even for short readout times, unlike the more traditional schemes affected by overlapping measurement results. It obtains good results also for longer measurement time where GMM method decrease its efficiency due to energy relaxation processes and a simple feed-forward neural network becomes difficult to train properly resulting in fluctuating results.

In addition, the proposed method allows for good classification on shorter measures, achieving a measurement speedup.

More importantly, this measurement speedup is helpful for real-time control systems, e.g., quantum orchestration platforms or quantum error correction, where we need to disturb the system as little as possible.

In general, it was shown that the proposed method performs well for all measurement times, helping in increasing classification results from a software point of view. On the other side, the classification times for a single measure are higher than standard methods but can be improved with more optimized FPGA and ASIC implementations. Lastly, the proposed approach can be readily extended to an arbitrary number of states (or, possibly, a number of qubits) with minimal modification of the model structure and obtaining marginally increasing performances.


\section{Acknowledgments}
This research was partially supported by Q@TN grants ML-QForge (PL). The Lawrence Livermore work was performed under the auspices of the U.S. Department of Energy by Lawrence Livermore National Laboratory under Contract DE-AC52-07NA27344 with support from Laboratory Directed Research and Development grant 19-DR-005 LLNL-JRNL-842516. 
P.E.T. acknowledges the Q@TN consortium for his support.
AR is funded by the European Union under Horizon Europe Programme - Grant Agreement 101080086 — NeQST. Views and opinions expressed are however those of the author(s) only
and do not necessarily reflect those of CERN, the European Union or European Climate, Infrastructure and Environment Executive Agency (CINEA). Neither CERN, the European Union nor the granting authority can be held responsible for them.

\appendix
\section{Numerical Consideration on Autoencoder}\label{sec:appendix}

\subsection{Autoencoder's latent space dimension}\label{subsec:appA}

\begin{figure}[ht]
\centering
\includegraphics[width=1\linewidth]{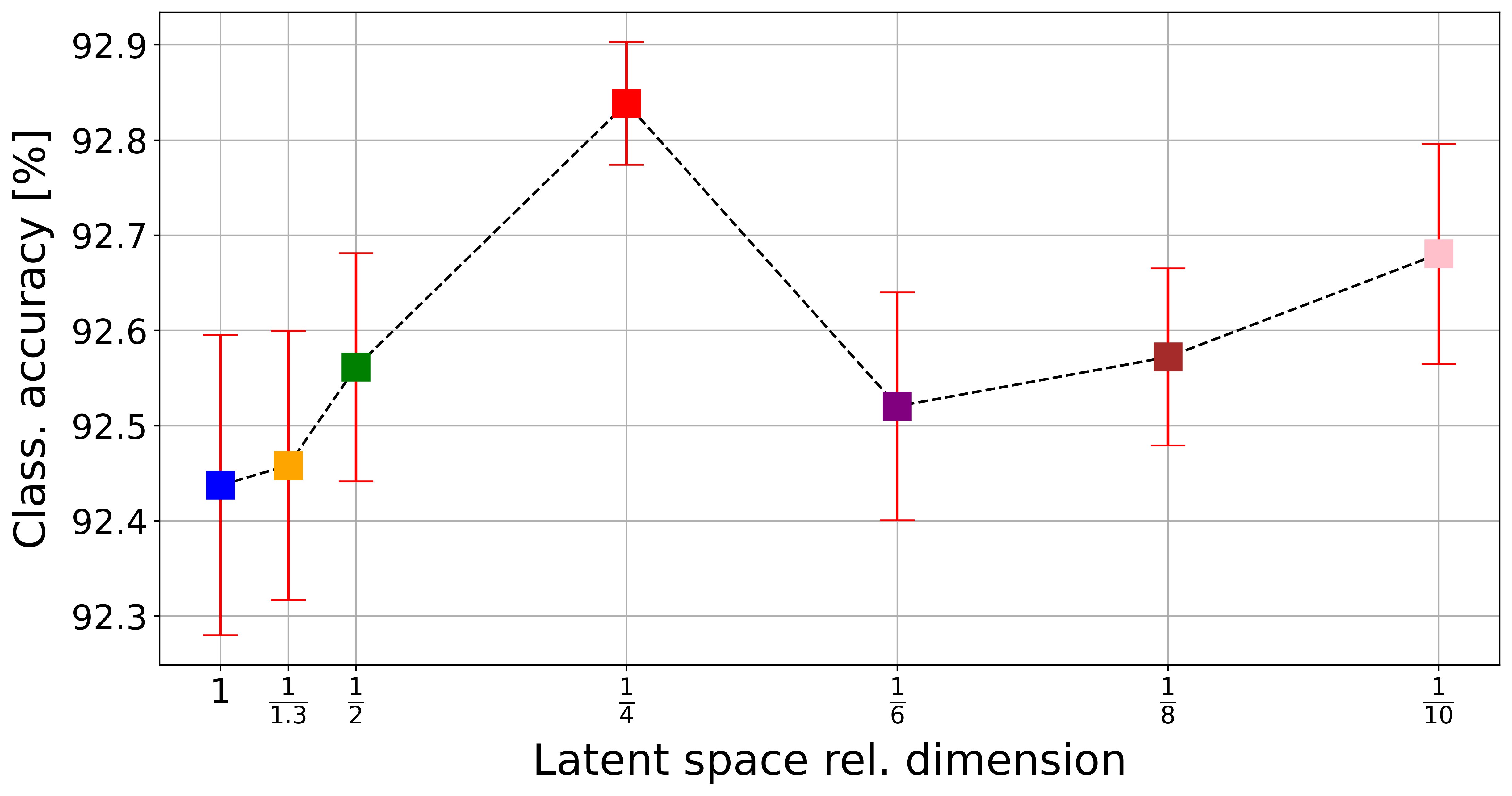}
\caption{PreTraNN global classification accuracy for the 3-state case with 2400 ns readout inputs as a function of the latent space dimension. The higher accuracy is reached at 1/4 the input dimension.}
\label{fig:accuracy_dim_analysis}
\end{figure} 

In the design of the architecture of a neural network, there is no solid theoretical guidance but one has to rely on a heuristic and "trial and error" attitude based on experience. However, to make the procedure more quantitative, one can vary the structure in an automated way and study how its metrics vary. In this way, one can identify, within a certain degree of approximation, the architecture that works best for the specific problem.

\begin{figure}[b]
\centering
\includegraphics[width=1\linewidth]{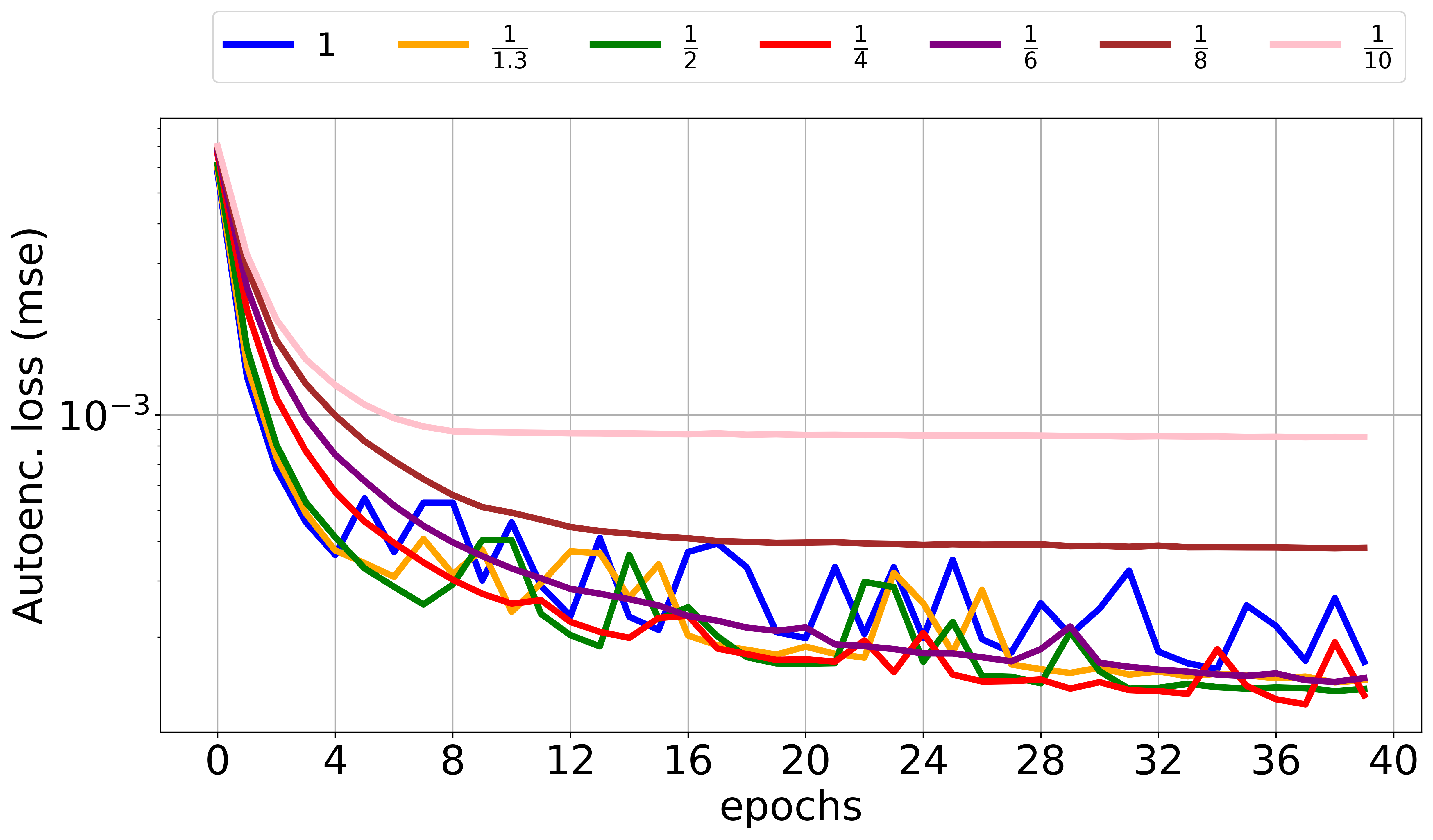}
\caption{Autoencoder training loss function as a function of training epochs for different latent space relative dimension. To large latent dimensions (1, 1/1.3 1/2 times the input size) present a fluctuating behavior and are useless for feature extraction while too small latent dimensions do not allow an effective encoding and their loss function remains high (1/8 and 1/10 the input size). }
\label{fig:loss_dim_analysis}
\end{figure}

\begin{figure}[ht]
\centering
\includegraphics[width=1\linewidth]{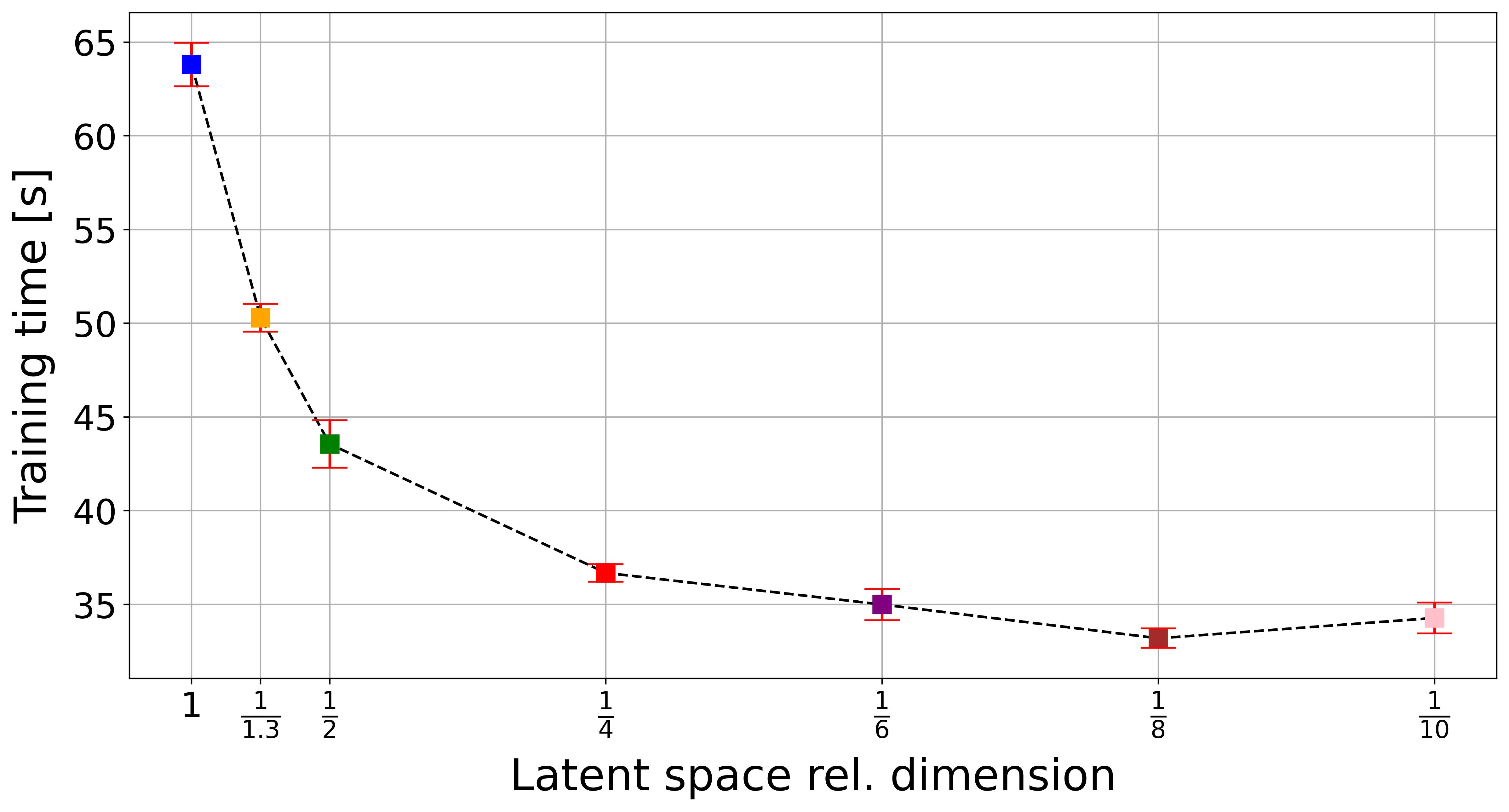}
\caption{Autoencoder training time in function of latent space relative dimension. Clearly larger latent spaces correspond to neural networks with more parameters and thus longer training times.}
\label{fig:train_time_dim_analysis}
\end{figure} 

In the case of the autoencoder, the main parameter is the autoencoder's latent space size. In principle, a latent space that is too small is not sufficient to perform expressive encoding, while too large of a latent space increases the computational cost without extracting in a compact way information from the dataset. In the limiting case of a latent space equal to the input space, the neural network becomes equivalent to applying an identity to the inputs.

In this appendix, we describe the procedure used in our work to identify the best autoencoder structure. 
We took the PreTraNN with trajectories of 2400 ns (150 time-steps of 16 ns, i.e. inputs dimension of 300 values), and trained it for different values of latent space. We started from a latent dimension equal to the input dimension and gradually went down to one-tenth of it. The dimension of the other two inner layers was set linearly interpolating between the size of the input and latent space. The decoder had the same structure but reversed. Contextually, three properties of PreTraNN were studied as a function of latent dimension: the global classification accuracy, the autoencoder training loss and autoencoder training time. To obtain more consistent results, for each latent dimension the training was repeated 10 times with different samplings of the dataset and the properties values was averaged.

In Fig. \ref{fig:accuracy_dim_analysis}
the PreTraNN global classification accuracy for decreasing latent space dimension is reported. The abscissa shows the size of the latent space in terms of fractions of the input length (so that the information extracted from this case can be scaled directly to the other input lengths). The greatest accuracy, moreover with the smallest error bars, is achieved with a latent space whose size is one-fourth that of the input space. In absolute terms, the classification accuracy is quite stable for every latent space dimension but an increasing trend from 1 to 1/4 can be clearly spotted.

In Fig.\ref{fig:loss_dim_analysis} is represented the loss function values (mean squared error) during the training of the autoencoder for different latent space dimensions. For large latent space sizes, the training converges faster for the first epochs but then assumes a fluctuating trend. For latent spaces that are small (e.g. 1/10, 1/8 the size of the input), on the other hand, convergence stalls at much higher values of the loss function. Thus the best values are 1/2, 1/4 and 1/6 of the input length.

In Fig. \ref{fig:train_time_dim_analysis} the training time in seconds is reported. Clearly, the training time decreases as the latent space decreases, since the number of network parameters decreases. A short training time is preferable.

Given this PreTraNN behavior, we can choose the latent space dimension making a trade-off between the reported metrics. The value which maximizes the classification accuracy having at the same time good loss function convergence and (relatively) short training time is a latent dimension of 1/4 the inputs size. This is the value chosen to carry out the analysis in this work. The dimension of the 2 internal layers is set linearly interpolating between the latent space and the input dimensions.

\subsection{Dataset size and convergence}\label{subsec:appB}

\begin{figure}[ht]
\centering
\includegraphics[width=1\linewidth]{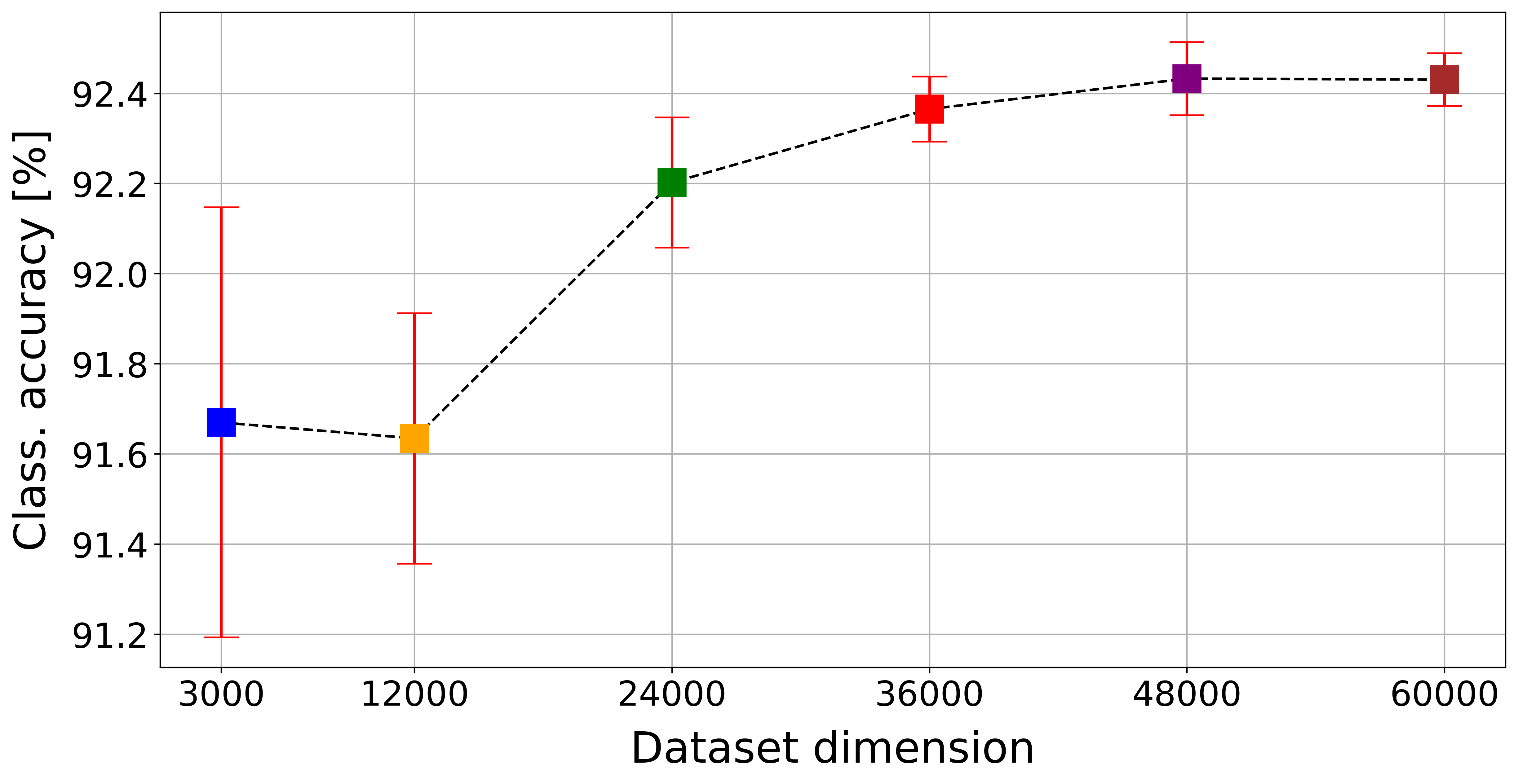}
\caption{Global classification accuracy of the PreTraNN as a function of the number of dataset elements.  }
\label{fig:accuracy_satur_dataset}
\end{figure} 

\begin{figure}[ht]
\centering
\includegraphics[width=1\linewidth]{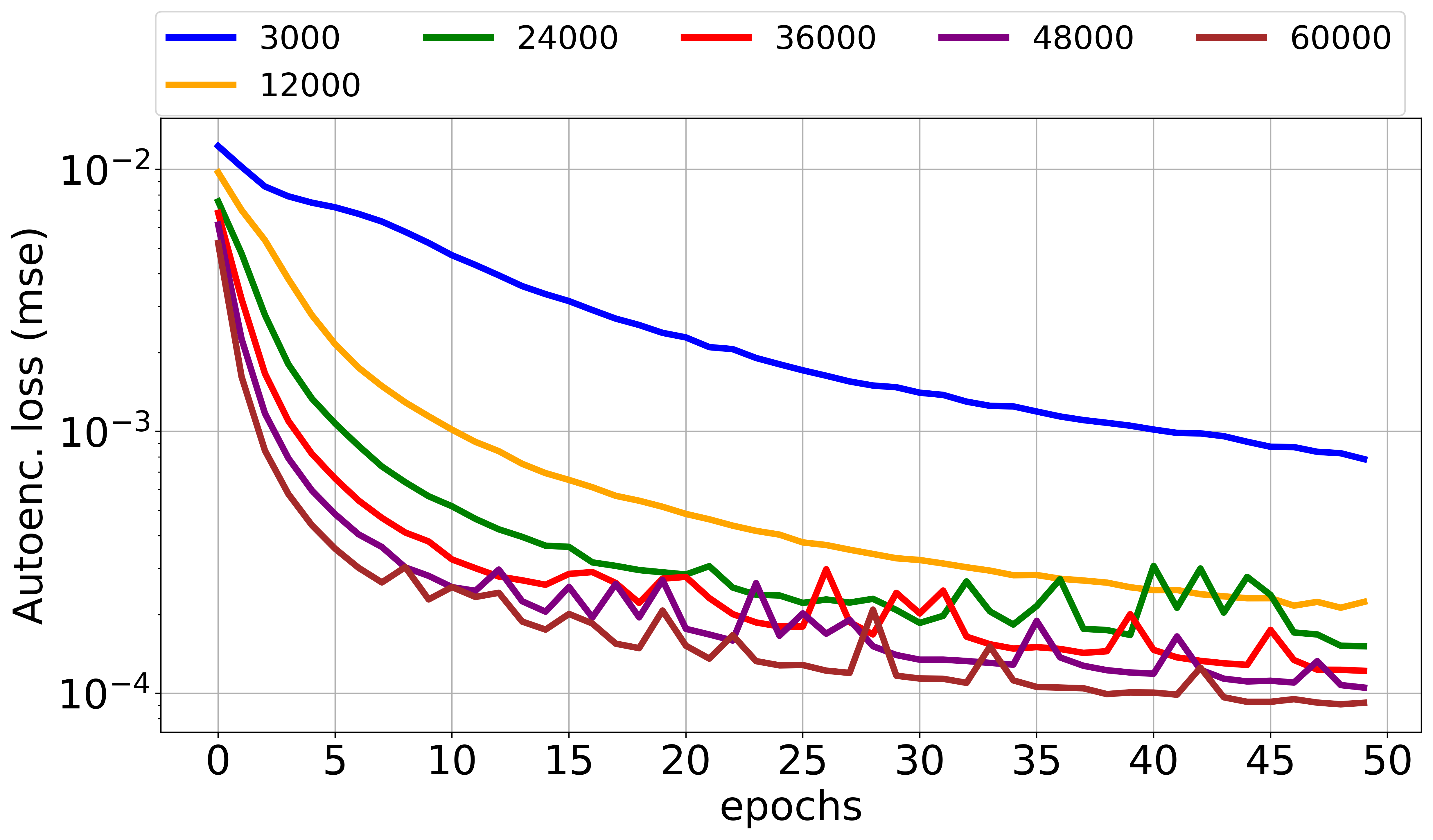}
\caption{Autoencoder loss (mean square error) as a function of the epochs for increasing dataset size.  }
\label{fig:loss_satur_dataset}
\end{figure} 

In order to obtain a good training convergence that maximizes classification accuracy an adequate dataset is needed. Small datasets are fast to train but usually produce inadequate classification accuracies, while large ones have the opposite behavior. At the same time, the growth of the classification accuracy capability is marginally decreasing with increasing dataset size. 
Here we report some analysis on the behavior of the PreTraNN as a function of the dataset dimension studying the same three properties introduced in the previous section i.e. loss function, classification accuracy and training time. Even in this case we took the PreTraNN with 2400 ns measurement signals (150 time-steps of 16 ns, i.e. inputs dimension of 300 values) with a latent space of 75 neurons, and trained it for different dataset dimensions. We started from a training dataset of 3000 elements (1000 elements for each class) and gradually increase its dimension to 60000 elements (with 75\% of them dedicated to training). For each dataset dimension, the training was repeated 10 times with different sampling of the dataset and the properties values were averaged.

In Fig.  \ref{fig:accuracy_satur_dataset} the global classification accuracy as a function of the dataset size is reported. It can be seen that the accuracy increases as the dataset grows even if with decreasing speed.

\begin{figure}[ht]
\centering
\includegraphics[width=1\linewidth]{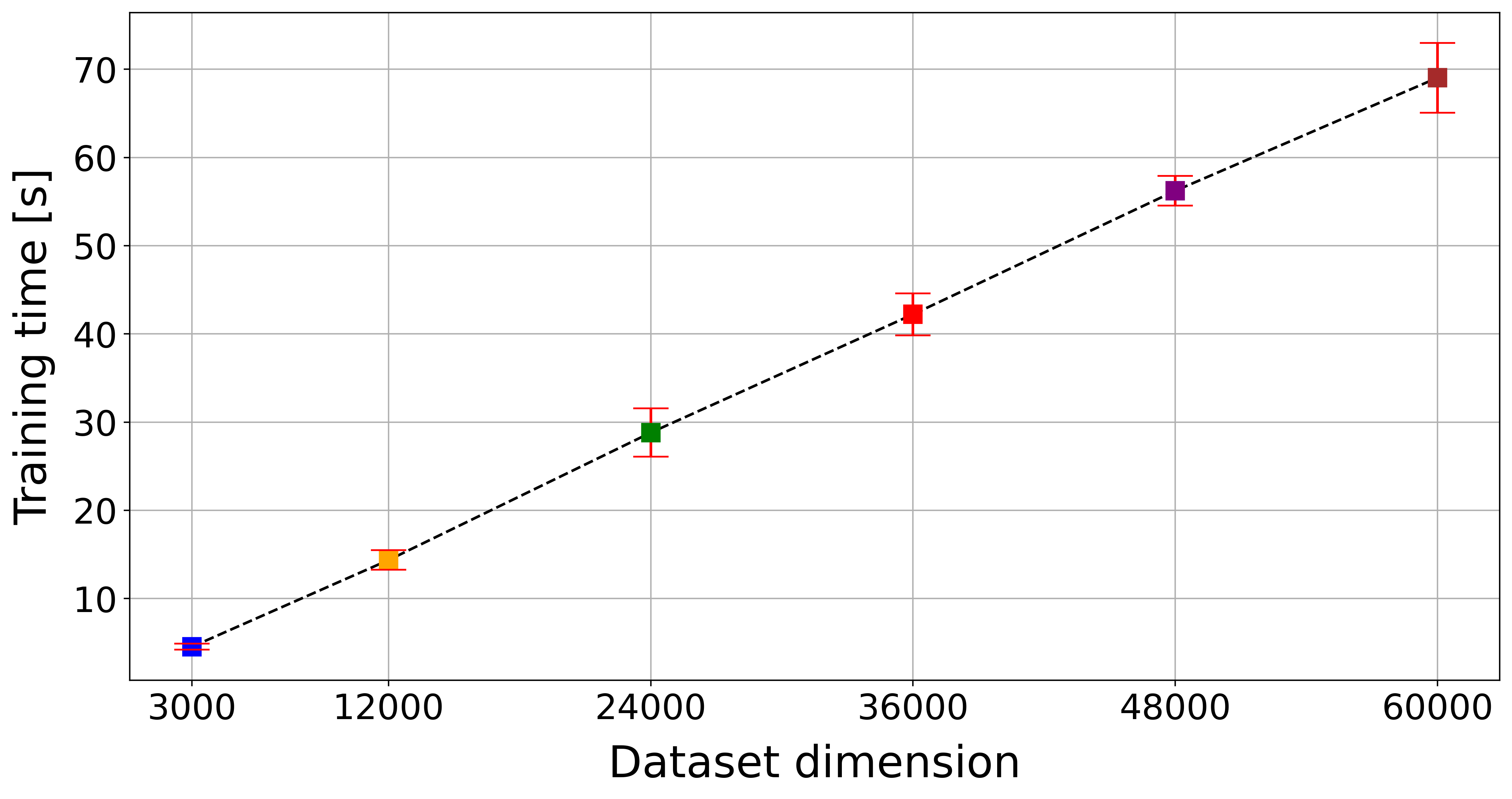}
\caption{Training time for increasing dataset dimension.}
\label{fig:train_time_satur_dataset}
\end{figure}

Fig. \ref{fig:loss_satur_dataset} represents the loss function values (mean squared error) during the training of the autoencoder for different configurations. The trend is quite neat. The larger the dataset the better the convergence, although for large data sets the convergence becomes more unstable.

In Fig. \ref{fig:train_time_satur_dataset} the training time in seconds is reported. As expected, the training time increase linearly with the dataset dimension. A short training time is preferable.

Given these results, the trade-off between accuracy, loss function, and training time, in order to maximize effectiveness and minimize cost, was identified in the 24000-item dataset for the three-state case  and the 16000-item dataset for two-state case.

\section{Models specifications}
\label{sec:appendixB}
We report here the complete characterization of the autoencoder, the PreTraNN, the FFNN and the GMM models and their procedure of training.

In this work, the building and training of the neural network are performed via the python package \emph{Keras} \cite{keras}. For the GMM instead the \emph{sklearn} python package \cite{sklearn}. 

\paragraph{Autoencoder} In every configuration employed in this work, the encoder is composed of an input layer, a first hidden layer and a second hidden layer connected to the latent layer. The decoder, on the other hand, has the same structure but is mirrored. So it has a first hidden layer connected to the latent layer, a second hidden layer and finally an output layer. We employ a full connectivity network implemented with the \emph{Dense} layer specification in Keras. In Tab. \ref{tab:autoencoder_spec} all the information on the network is reported.

The training is performed using the \emph{Adam} stochastic optimization algorithm \cite{kingma2014adam} with the standard configuration implemented in Keras.
The loss function is the \emph{mean square error}. The training is performed with the  \emph{Early Stopping} procedure that stops the training if the loss does not decrease for two epochs in a row.

\begin{table}
\begin{tabular}{l|l|l|l|l|}
 & Layer & Size & \begin{tabular}[c]{@{}l@{}}Activ.\\ funct.\end{tabular} & \begin{tabular}[c]{@{}l@{}}Keras\\ type\end{tabular} \\ \hline
\multirow{3}{*}{Encoder} & input & L & sigmoid & Dense \\ \cline{2-5} 
 & $1^{th}$ hidden & L3/4 & tanh & Dense \\ \cline{2-5} 
 & $2^{nd}$ hidden & L2/4 & tanh & Dense \\ \hline
 & latent & L/4 & tanh & Dense \\ \hline
\multirow{3}{*}{Decoder} & $1^{th}$ hidden & L2/4 & tanh & Dense \\ \cline{2-5} 
 & $2^{nd}$ hidden & L3/4 & tanh & Dense \\ \cline{2-5} 
 & output & L & sigmoid & Dense \\ \hline
\end{tabular}
\caption{Autoencoder's specifications. The "Size" column represents the number of neurons for each layer in a fraction of the input dimension $L$. The "Keras type" column reports the type of Keras layer employed.  \label{tab:autoencoder_spec}}
\end{table}

\paragraph{ FFNN and PreTraNN's second stage} The second stage of the PreTraNN is a simple feed-forward neural network. It is composed of an input layer (of the same dimension as the latent layer of the autoencoder), a first hidden layer and a second hidden layer connected to the output layer. The dimension $C$ of the output layer depends on the number of classes we are doing the classification with. Hence, $C=2$ for qubit classification of Sec. \ref{subsec:3A}, while $C=3$ for qutrit classification of Sec. \ref{subsec:3B}. The connectivity between the neurons is full. The optimization algorithm is the \emph{Adam}. The loss function is the \emph{cross-entropy}, suitable for classification purposes. The training is performed with the \emph{Early Stopping} procedure that stops the training if the loss does not decrease for two epochs in a row. Other information is summarized in Tab. \ref{tab:FFNN}. The structure of FFNN model is the same but with a number of input neurons equal to the dataset dimension instead of the latent layer dimension.

\begin{table}[]
\begin{tabular}{l|l|l|l|}
Layer & Size & \begin{tabular}[c]{@{}l@{}}Activ.\\ funct.\end{tabular} & \begin{tabular}[c]{@{}l@{}}Keras\\ type\end{tabular} \\ \hline
Input & L/4 (L) & tanh & Dense \\ \hline
$1^{th}$ hidden & L2/4 (2L) & tanh & Dense \\ \hline
$2^{nd}$ hidden & L/4 (L) & tanh & Dense \\ \hline
Output & C & softmax & Dense \\ \hline
\end{tabular}
\caption{Structure and specifications of PreTraNN's second section (FFNN) network with Keras. $L$ is the dataset inputs length, $C$ is the dimension of the output layer which change based on the number of classes. \label{tab:FFNN} }
\end{table}

\paragraph{Gaussian Mixture Model} The GMM is implemented with \emph{sklearn} package with the standard build-in parameters specifying only the number of classes of the input dataset.

\section{Autoencoder features}\label{sec:appendixC}

\begin{figure}[ht]
\centering
\includegraphics[width=0.8\linewidth]{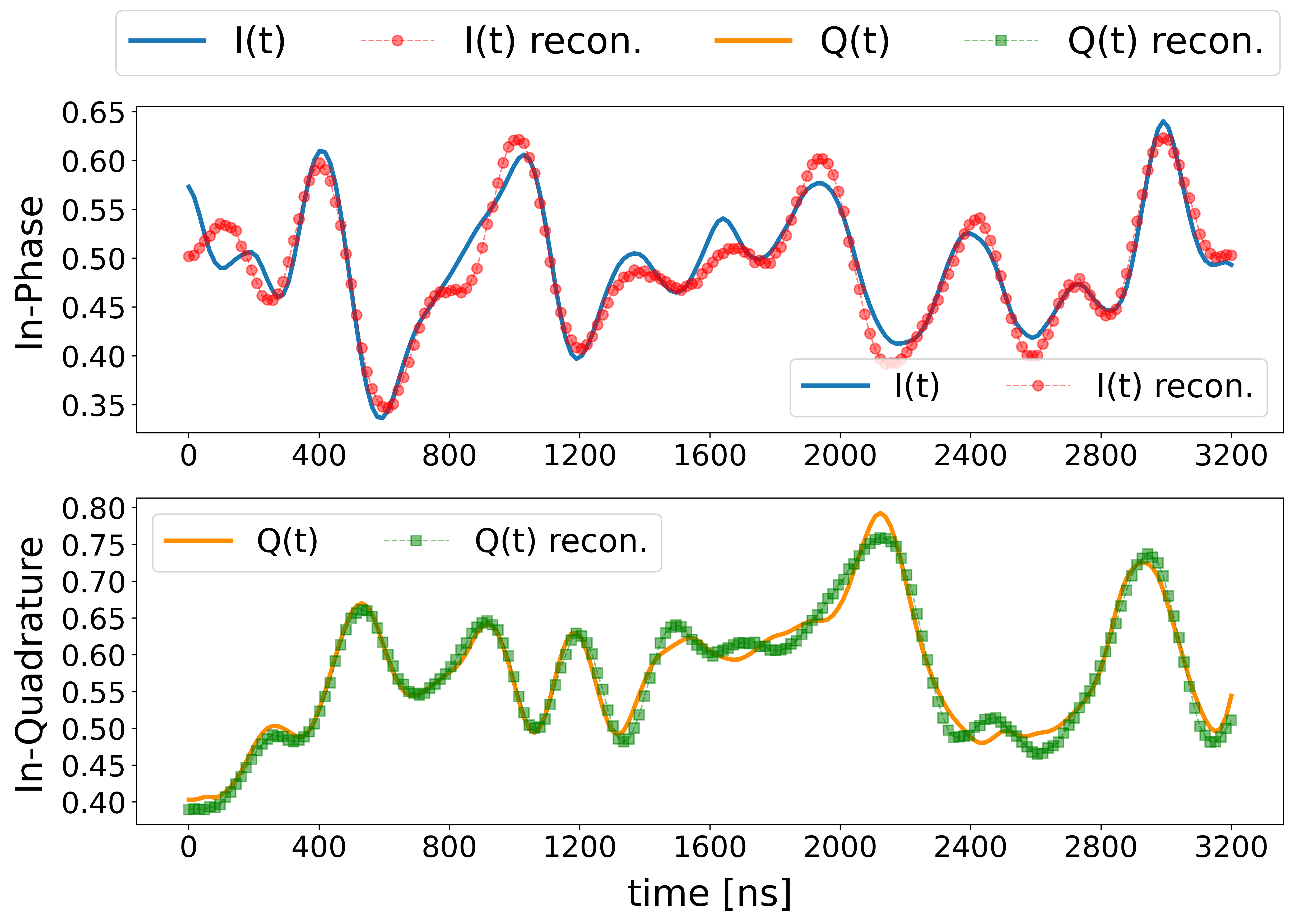}
\caption{An example of input regeneration made by the autoencoder. In both panels, the solid lines represent the measurement signal divided into its two quadratures, respectively In-phase (I) and In-Quadrature (Q). The lines with markers, instead, represent the input reconstruction made by the autoencoder.}
\label{fig:real_recon_compar_01}
\end{figure}

In this Appendix, we give examples of the two important autoencoder features: input regeneration and latent space values. 
Fig.~\ref{fig:real_recon_compar_01} shows an example of  3200 ns (i.e. 400 components) input reconstruction done by the autoencoder. The solid lines represent the original input (divided into the two quadratures), while the lines with markers represent the output of the autoencoder, i.e., the regeneration of the input from its synthetic representation in the latent space of the autoencoder. It can be seen that the reconstruction is quite faithful to the original.

\begin{figure}[ht]
\centering
\includegraphics[width=0.8\linewidth]{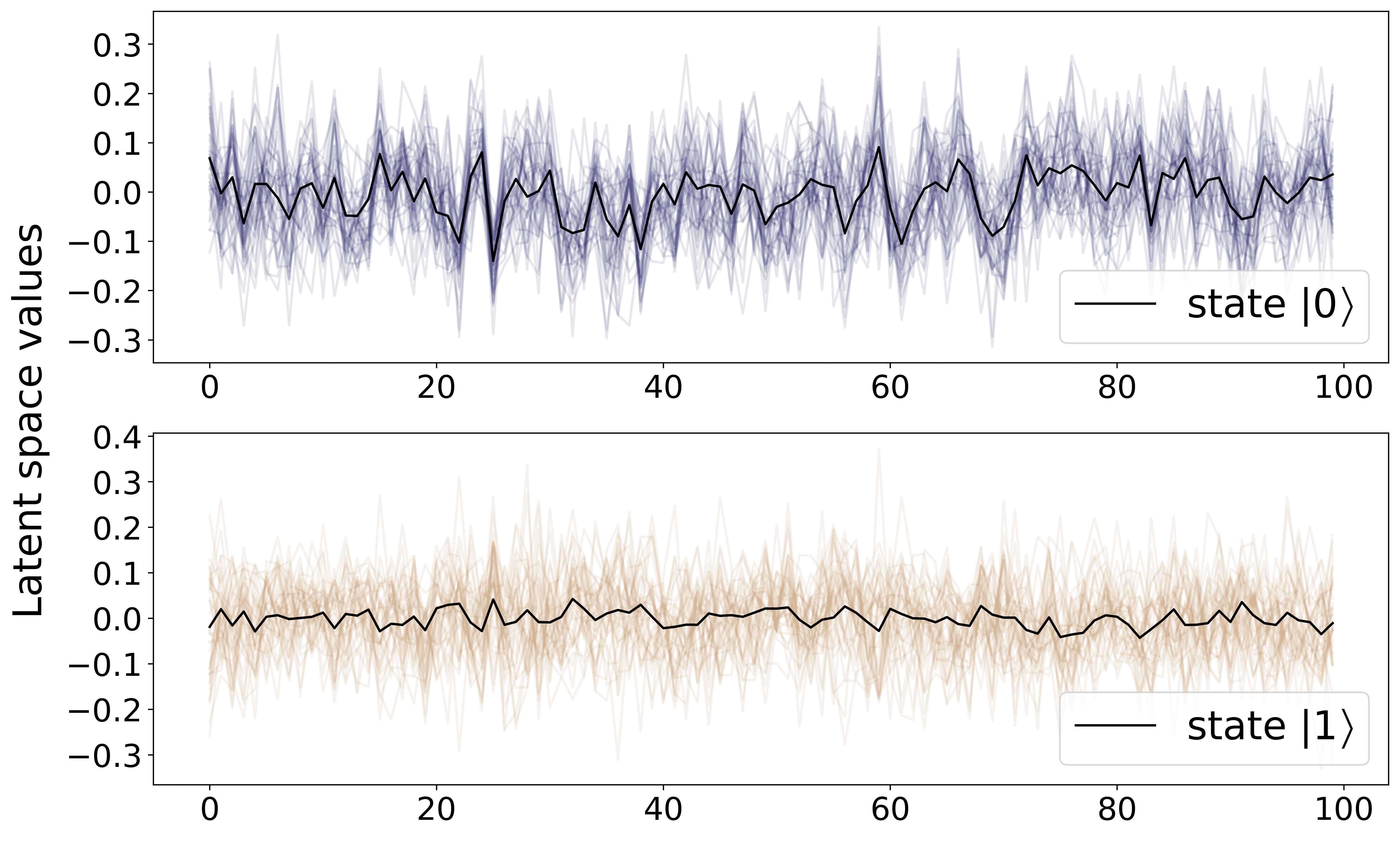}
\caption{Representation of latent space of the autoencoder for state $\ket{0}$ (upper) and $\ket{1}$ state (lower). In both panels, the colored lines are the latent space representation (i.e. $\bm h^i$ vector) of inputs for state $\ket{0}$ or $\ket{1}$. The solid black lines represent instead the average of these values.}
\label{fig:latent_space_01}
\end{figure}

The latent space representation is presented in Fig. \ref{fig:latent_space_01}. The thin colored lines represent the latent space values of different inputs while the thick black line is the average of such lines. It can be seen that the latent space vectors for the two states are somewhat different on average. Both have 0 on average but those for $\ket{0}$ have larger fluctuations and a bit of structure. In particular, in both plots specific points where all the $\bm h^i$ vectors follow a definite trend (e.g., the points around 20 and 60 for state $\ket{0}$ ) can be spotted. These differences are the ones that allow the increase in classification performance shown in this paper.

\begin{figure}[ht]
\centering
\includegraphics[width=0.8\linewidth]{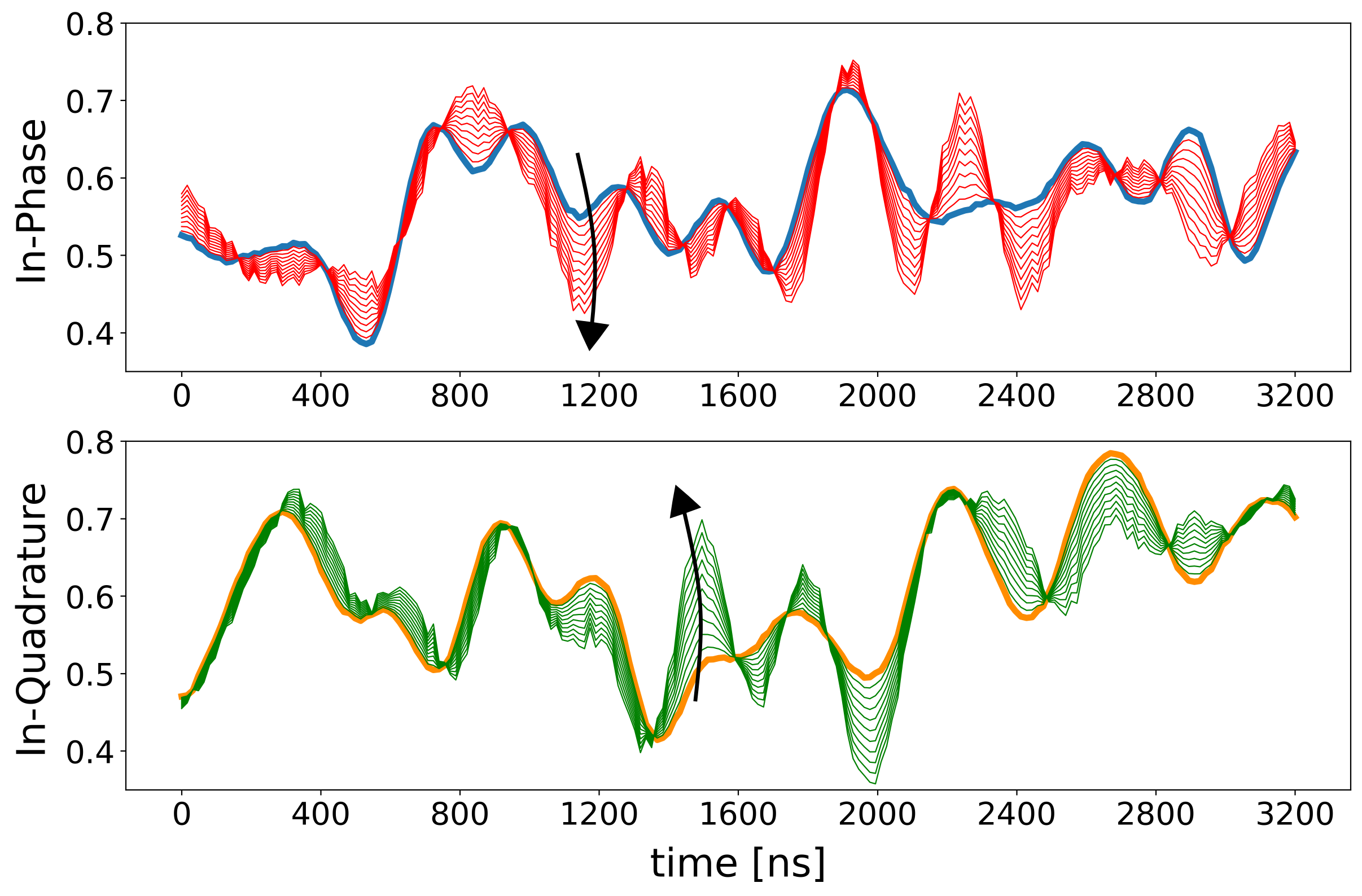}
\caption{The figure depicts an example of how the input reconstruction varies if a single value of the latent representation is varied slightly. The upper panel represents the in-phase component, lower panel the quadrature one. In both panels, the thick lines are the original "correct" input reconstruction, and the thin lines represent the reconstructions obtained by slowly varying a single value of the latent representation. In both panels, arrows are used to indicate the direction of changes induced by increasing the latent value. }
\label{fig:latent_space_variation_01}
\end{figure}

One might wonder how inputs reconstruction varies as the latent representation varies. To answer this question we can proceed as follows.  We use the encoder to obtain the latent representation of an input, we then vary slightly only one of its values, and finally, we plug the modified latent vector into the decoder to obtain its "reconstruction". We do this several times by varying slightly the input each time. Fig.~\ref{fig:latent_space_variation_01} depicts the result of this procedure. The thick lines represented the correct reconstruction of an input (divided into I and Q components) while the thin lines represent the reconstruction for increasing values of the 20th component of the latent representation. We can see that by slowly varying this value, we obtain a slowly varying family of reconstructions.

\section{Other classification data}\label{sec:appendixD}
In this appendix, we report classification accuracy data for the single states of the three-level qutrit case introduced in Sec. \ref{subsec:3B}.

\begin{figure}[htbp]
  \centering
  \caption{State-by-state classification accuracy for the qutrit case.}
  \begin{subfigure}[b]{0.37\textwidth}
    \includegraphics[width=\textwidth]{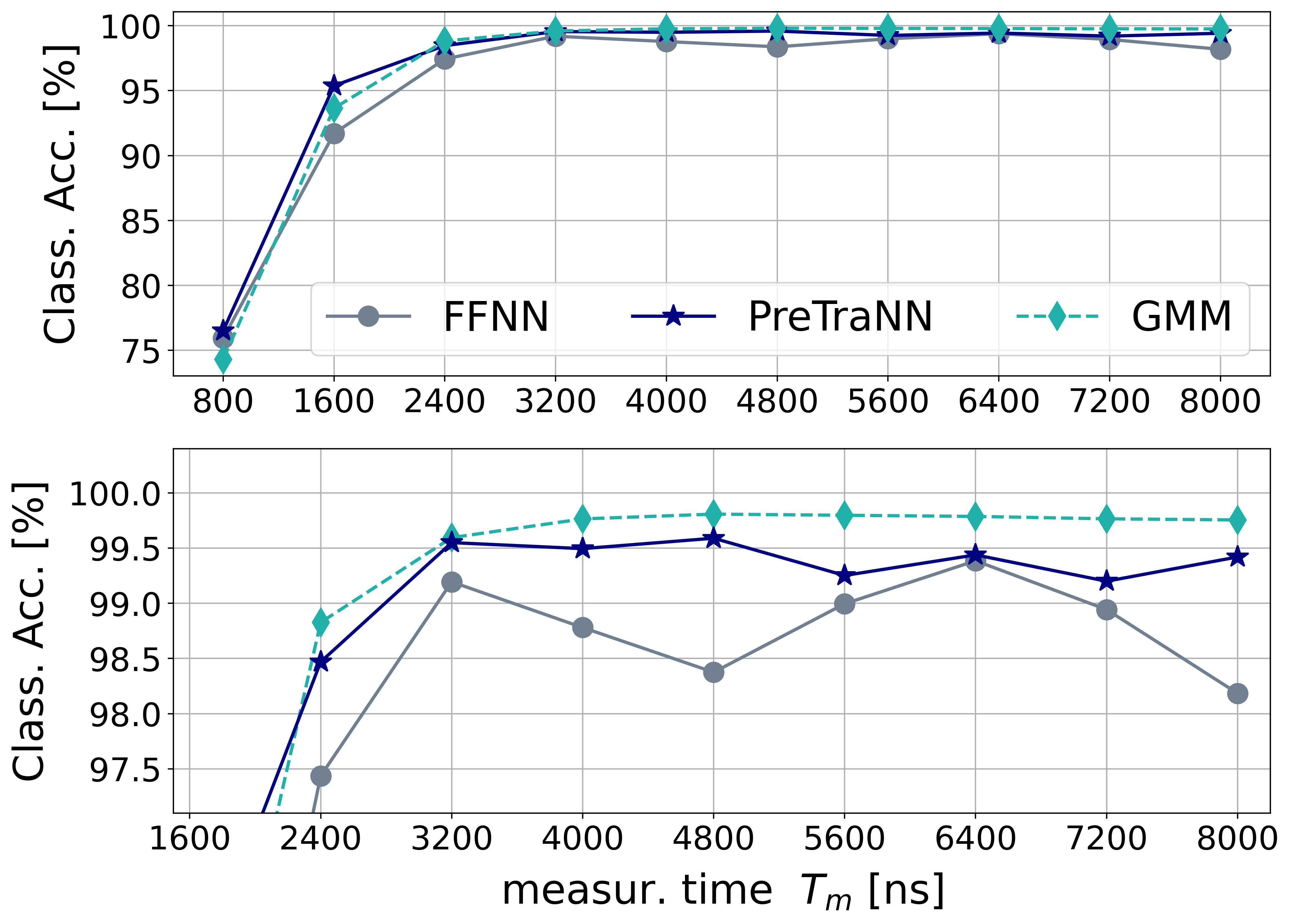}
    \caption{\emph{Upper panel}: State $\ket{0}$ classification accuracy for the three methods as a function of the measurement time in the case of a qutrit. \emph{Lower panel}: zoom on the medium-long times.}
\label{fig:accuracy_0_012}
  \end{subfigure}
  \hfill
  \begin{subfigure}[b]{0.37\textwidth}
    \includegraphics[width=\textwidth]{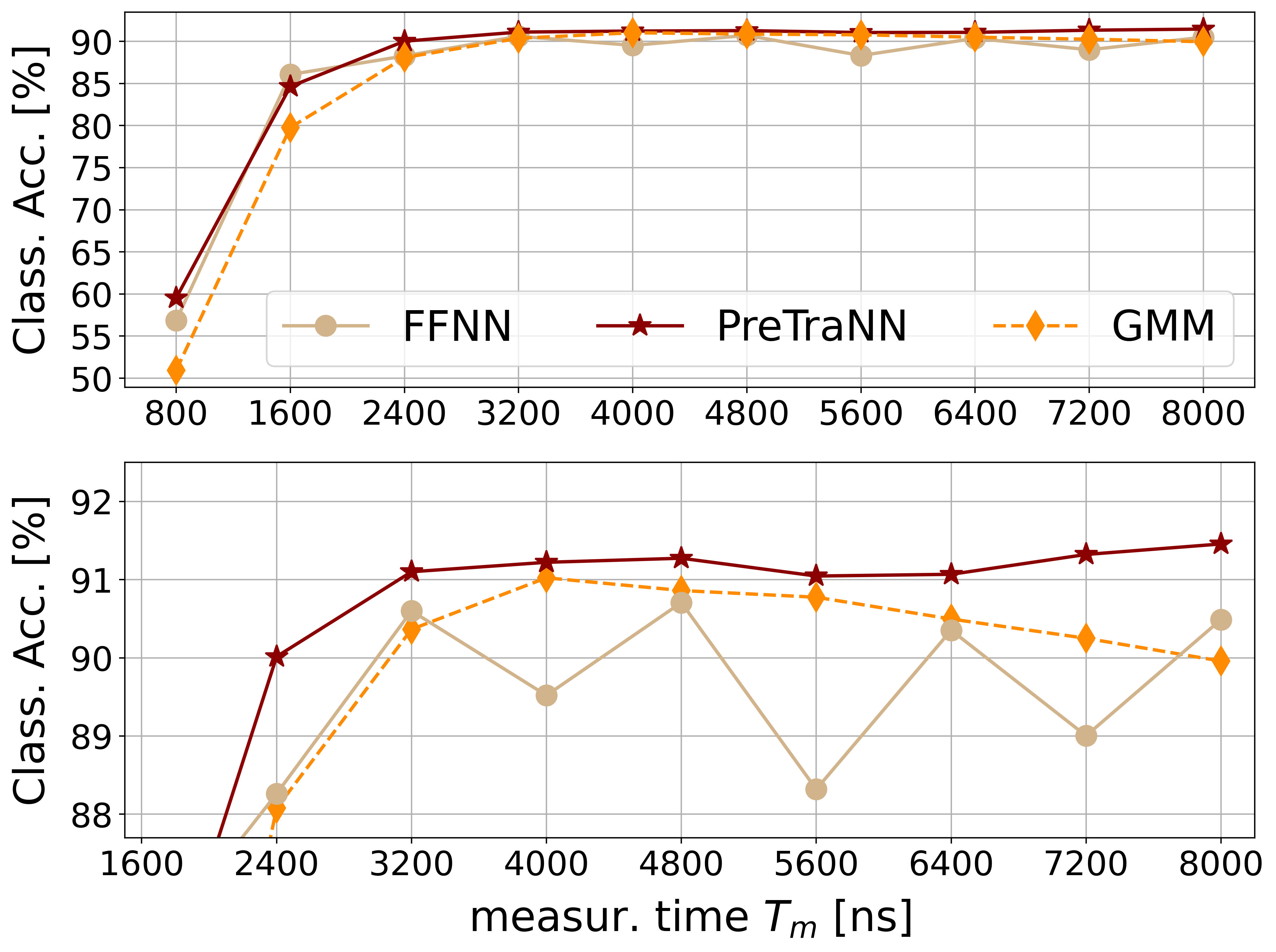}
    \caption{\emph{Upper panel}: State $\ket{1}$ classification accuracy for the three methods as a function of the measurement time in the case of a qutrit. \emph{Lower panel}: zoom on the medium-long times.}
\label{fig:accuracy_1_012}
  \end{subfigure}
  \hfill
  \begin{subfigure}[b]{0.37\textwidth}
    \includegraphics[width=\textwidth]{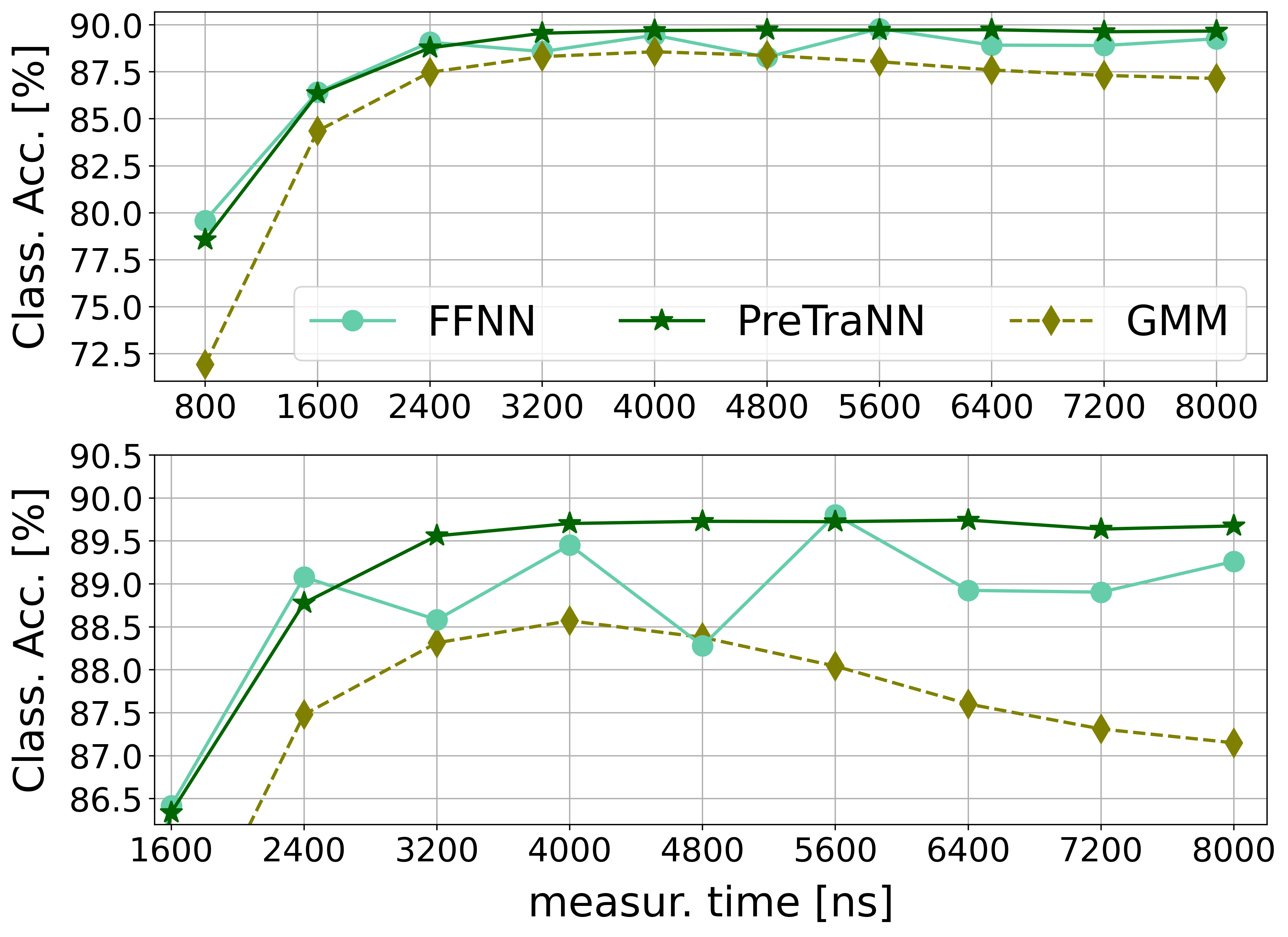}
    \caption{\emph{Upper panel}: State $\ket{2}$ classification accuracy for the three methods as a function of the measurement time. \emph{Lower panel}: zoom on the medium-long times.}
\label{fig:accuracy_2_012}
  \end{subfigure}
\end{figure}

In Fig.~\ref{fig:accuracy_0_012},\ref{fig:accuracy_1_012},\ref{fig:accuracy_2_012} we show the classification accuracy for, respectively, state $\ket{0}$,$\ket{1}$ and $\ket{2}$. The lower panel of each figure is a zoom on the 2400-8000 ns part of the plot to better see the details. Even in this configuration, we can see the same trends as in the 2-level case. All methods show bad results for short times, especially GMM, and the FFNN still exhibits a seesaw pattern that makes it poorly suited to the task. Again, GMM performs better than PreTraNN in state $\ket{0}$ and worse in state $\ket{1}$ classification due to the data distribution asymmetry. For state $\ket{2}$ the difference between GMM and PrTranNN is even higher since the state $\ket{2}$ can decay not only on the state $\ket{0}$ but also on state $\ket{1}$.

\newpage

\bibliography{bibliography.bib}

\end{document}